\newif\ifreview

\documentclass[sigconf]{acmart}

\usepackage[fleqn]{mathtools}
\usepackage{xspace}

\newcommand{\system}{\textsc{LitPivot}\xspace}

\definecolor{revcolor}{RGB}{128,0,128}
\reviewfalse
\ifreview
  \newcommand{\hita}[1]{\textcolor{revcolor}{#1}}
  \newenvironment{hitapar}{\color{revcolor}}{}
\else
  \newcommand{\hita}[1]{#1}
  
\fi
\newcommand{\hitacut}[1]{}

\AtBeginDocument{%
  }

\copyrightyear{2026}
\acmYear{2026}
\setcopyright{cc}
\setcctype{by}
\acmConference[UIST '26]{The 39th Annual ACM Symposium on User Interface Software and Technology}{November 02--05, 2026}{Detroit, MI, USA}
\acmBooktitle{The 39th Annual ACM Symposium on User Interface Software and Technology (UIST '26), November 02--05, 2026, Detroit, MI, USA}
\acmDOI{10.1145/3830398.3830577}
\acmISBN{979-8-4007-2856-3/2026/11}

\begin{document}
\title[LitPivot]{LitPivot: Developing Well-Situated Research Ideas Through Dynamic Contextualization and Critique \\ within the Literature Landscape}



\begin{abstract}

Developing a novel research idea is hard. It must be distinct enough from prior work to claim a contribution while also building on it. This requires iteratively reviewing literature and refining an idea based on what a researcher reads; yet when an idea changes, the literature that matters often changes with it. Most tools offer limited support for this interplay: literature tools help researchers understand a fixed body of work, while ideation tools evaluate ideas against a static, pre-curated set of papers. We introduce \textit{literature-initiated pivots}, a mechanism where engagement with literature prompts revision to a developing idea, and where that revision changes which literature is relevant. We operationalize this in \system{}, where researchers concurrently draft and vet an idea. \system{} dynamically retrieves clusters of papers relevant to a selected part of the idea and proposes literature-informed critiques for how to revise it. A user study ($n{=}17$) shows researchers produced ideas rated by experts as better grounded, perceived their ideas as more novel and useful, and reported stronger understanding of the literature space. An open-ended study ($n{=}5$) reveals how researchers use \system{} to iteratively evolve their own ideas.
\end{abstract}

\author{Hita Kambhamettu}
\authornote{Work completed during an internship at Ai2.}
\orcid{0000-0001-9620-1533}
\email{hitakam@seas.upenn.edu}
\affiliation{%
  \institution{University of Pennsylvania}
  \city{Philadelphia}
  \country{USA}
}

\author{Bhavana Dalvi Mishra}
\authornote{Now at Google; work conducted while at Ai2.}
\orcid{0000-0002-3813-8641}
\email{bh3avana.dalvi2023@gmail.com}
\affiliation{%
  \institution{Allen Institute for AI}
  \city{Seattle}
  \country{USA}
}

\author{Andrew Head}
\orcid{0000-0002-1523-3347}
\email{head@seas.upenn.edu}
\affiliation{%
  \institution{University of Pennsylvania}
  \city{Philadelphia}
  \country{USA}
}

\author{Jonathan Bragg}
\orcid{0000-0001-5460-9047}
\email{jbragg@allenai.org}
\affiliation{%
  \institution{Allen Institute for AI}
  \city{Seattle}
  \country{USA}
}

\author{Aakanksha Naik}
\orcid{0000-0002-3673-0051}
\email{aakankshan@allenai.org}
\affiliation{%
  \institution{Allen Institute for AI}
  \city{Seattle}
  \country{USA}
}

\author{Joseph Chee Chang}
\orcid{0000-0002-0798-4351}
\email{josephc@allenai.org}
\affiliation{%
  \institution{Allen Institute for AI}
  \city{Seattle}
  \country{USA}
}

\author{Pao Siangliulue}
\orcid{0009-0006-8042-885X}
\email{paos@allenai.org}
\affiliation{%
  \institution{Allen Institute for AI}
  \city{Seattle}
  \country{USA}
}

\renewcommand{\shortauthors}{Kambhamettu et al.}



\begin{teaserfigure}
\centering
  \includegraphics[width=0.6\textwidth]{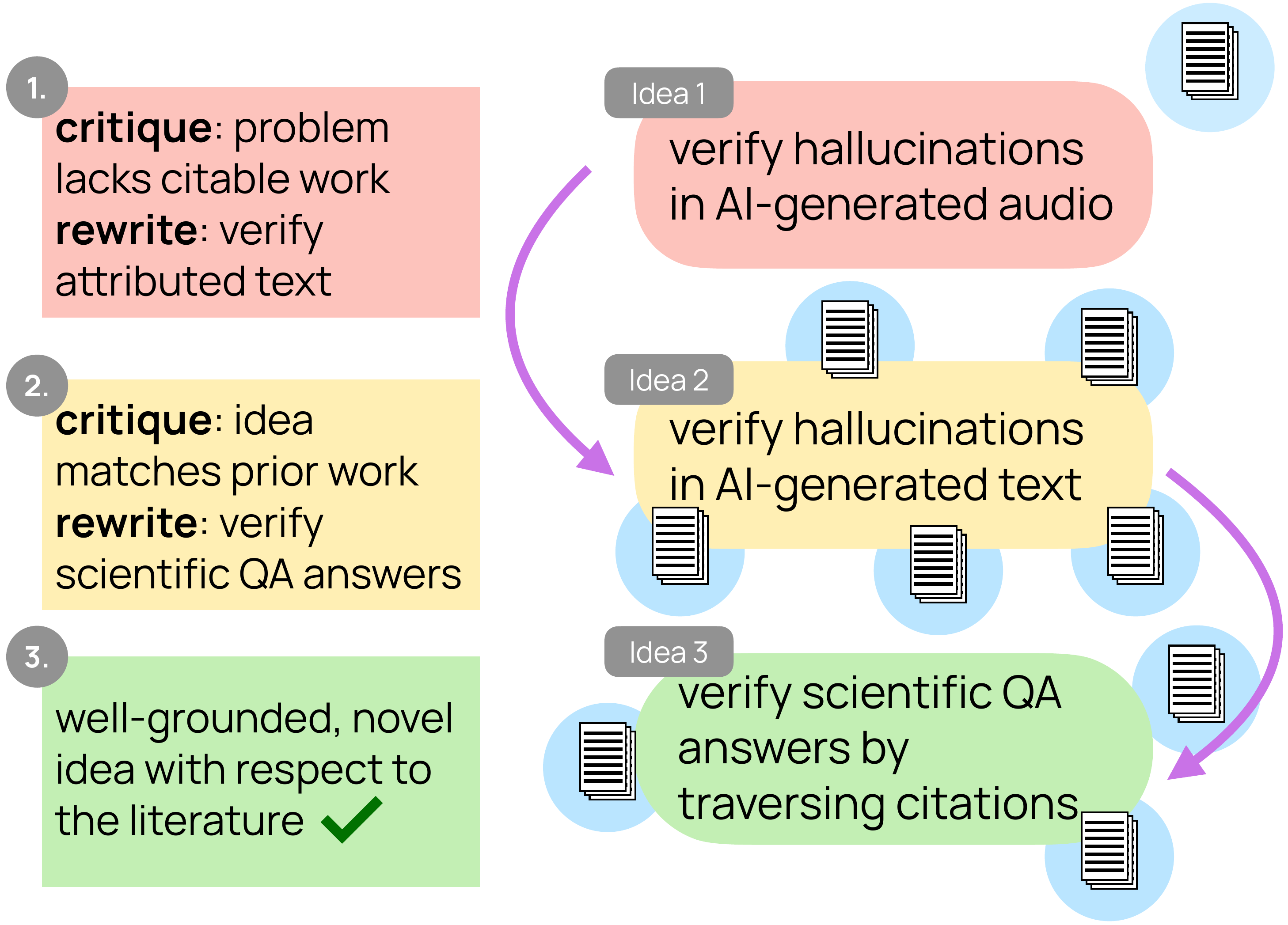}
  \caption{\normalfont \hita{\textbf{\system{} supports the co-development of a research idea and its grounding in the literature through \textit{literature-initiated pivots} (in purple).} (1) The authors' original idea (verifying hallucinations in AI-generated audio) has too little prior work to motivate it, so \system{} suggests pivoting to verifying hallucinations in AI-generated text. (2) That idea is not novel, so \system{} suggests traversing retrieved citations in scientific RAG systems. (3) The resulting idea builds on prior work while remaining differentiated from it.}}
  \Description{A diagram in two columns tracing three versions of a research idea. The right column shows three rounded boxes stacked vertically, labeled Idea 1, Idea 2, and Idea 3, each surrounded by small circular icons representing clusters of retrieved papers. Idea 1 reads "verify hallucinations in AI-generated audio" and has one nearby paper cluster. Idea 2 reads "verify hallucinations in AI-generated text" and is surrounded by five clusters. Idea 3 reads "verify scientific QA answers by traversing citations" and is bordered by three clusters. Purple curved arrows run from Idea 1 down to Idea 2, and from Idea 2 down to Idea 3, marking each literature-initiated pivot. The left column holds three numbered boxes aligned with the ideas. Box 1 reads "critique: problem lacks citable work" and "rewrite: verify attributed text." Box 2 reads "critique: idea matches prior work" and "rewrite: verify scientific QA answers." Box 3 reads "well-grounded, novel idea with respect to the literature" and ends with a check mark.}
  \label{fig:teaser}
\end{teaserfigure}


\maketitle

\section{Introduction}

This paper was originally going to be about a human-AI system to verify hallucinations in AI-generated audio files. How did we get here? After drafting an initial idea, we did a literature search and found that only two research groups were interested in this space. We were one of them. To ensure we addressed a problem important to the community, we pivoted to verifying hallucinations in AI-generated attributed text. A second literature search revealed the opposite problem: the space was saturated. While this validated the community's interest, it meant we would need to articulate a novel contribution. Our focus narrowed to helping users make sense of scientific RAG QA answers by traversing their retrieved academic citations. This raised a question: who needs to do that, and why? One answer was researchers developing an idea, determining whether existing literature supports it. In other words, us. Supporting that need, a researcher checking a developing idea against the literature, is ultimately what this paper contributes.

Changes to our idea were informed by literature. This is consistent with prior work that states that literature search and idea development are intertwined~\cite{diaz2022developing, finfgeld2013literature}: through literature review, researchers iteratively change their ideas as they compare against related work and reconsider what is novel, important, and feasible~\cite{granello2001promoting, knopf2006doing, lim2022advancing}. Existing literature can expose a flaw in an idea's motivation and inspire a stronger framing. The changes to the idea in turn shape which literature becomes most relevant. We call this phenomenon when engagement with literature prompts a revision to a developing idea, and where that revision changes which literature is relevant, a \textit{literature-initiated pivot}.

Prior theories of design and scientific reasoning also conceptualize phenomena similar to \textit{literature-initiated pivots}. Concept-Knowledge Theory describes design as movement between an evolving concept space and existing knowledge~\cite{hatchuel2003new}; related perspectives in scientific reasoning and creativity similarly emphasize reciprocal movement between generating possibilities and evaluating them against available knowledge~\cite{klahr1988dual, finke1996imagery}. These frameworks emphasize that the relevant knowledge landscape depends on how the hypothesis has evolved, and vice versa. These perspectives motivate a design opportunity for research ideation tools. Systems should support ideation as a tight, iterative loop: where actively developing an idea surfaces the literature that grounds, challenges, and reshapes it, which in turn meaningfully changes the idea.

Existing tools support understanding the literature space or developing the idea largely in isolation. On one hand, systems for augmented reading~\cite{head2021augmenting, August2022PaperPM, fok2023scim}, scholarly synthesis~\cite{kang2022augmenting, kang2022threddy}, and literature review~\cite{palani2023relatedly} help researchers read papers, identify themes across them, and build overviews of a research space. However, they provide less support for carrying insights from specific papers back into an evolving research idea~\cite{palani2023relatedly,kang2022augmenting}. 
On the other hand, ideation tools help researchers generate research questions~\cite{liu2024ai}, assess idea novelty~\cite{radensky2024scideator}, and iterate on an idea~\cite{pu2025ideasynth}. While some systems provide literature-grounded feedback~\cite{radensky2024scideator, pu2025ideasynth}, they use a fixed, user-curated set of papers assembled before ideation begins. 

In this work, we explore \textit{literature-initiated pivots} in the context of early-stage research ideation, where ideas evolve quickly and developing them requires reciprocal engagement with the literature~\cite{diaz2022developing}. We introduce \system, a system designed to support the co-development of a research idea and its grounding in a literature space. A researcher can check a facet of their idea (problem, contribution, or evaluation) against the literature that is retrieved and organized for that aspect. Using bipartite graph-based and LLM-simulated reasoning, \system maps the idea aspect onto the aspects of the retrieved literature, showing how the idea follows or differs from prior work and how the idea can be re-articulated to be more grounded or distinct. The researcher can then review this output, which includes references back to the literature, and integrate this updated understanding to refine their idea. \system\ is intended for researchers who already hold a nascent idea and want to develop it against the literature. \hita{While ideas can originate from many sources (e.g. conversations, observations, practical needs),  once an idea has been articulated, engaging with the literature is an important way researchers refine, differentiate, and strengthen it. This is the stage of ideation that \system\ supports.}

Three studies inform and validate \system. First, a formative study ($n=5$) and research artifact analysis ($n=4$) suggest challenges researchers face when making literature-initiated pivots. \hita{Second, in a user study ($n=17$) comparing \system to a baseline system that supports ideation in a popular current practice, AI chat-with-papers QA, we find that using \system leads researchers to consult significantly more papers when developing their idea, report a stronger understanding of the literature space, and produce ideas that were rated by experts as better grounded in the literature.} Finally, in a qualitative, open-ended study, we observe literature-initiated pivots in practice as participants used \system to develop their ongoing ideas in relation to the literature. 

\hita{We conclude by discussing how \system may have improved researchers' engagement with the literature and the grounding of their ideas, and how it shaped the ideas themselves}. We discuss how ideation in other knowledge-intensive domains may also benefit from interactive integration with a corpus of knowledge. Our work suggests that the corpus could be a more active participant in shaping new insights in these domains. In summary, this paper:  

\begin{itemize}
    \item Defines the \textit{literature-initiated pivot} as the process where engagement with literature prompts the revision of a developing idea, which in turn shifts the relevant literature, and identifies it as a key mechanism to support during ideation.
    \item Contributes a system, \system, that operationalizes support for literature-initiated pivots in early-stage research ideation through dynamic literature reorganization and graph-based reasoning over a literature space.
    \item \hita{Provides evidence from a user study ($n=17$) that supporting literature-initiated pivots leads researchers to report deeper understanding of the literature and develop better-grounded ideas. A qualitative study ($n=5$) illustrates how researchers benefit from making these pivots.}
\end{itemize}
\section{Related Work}

\subsection{Ideation as Dual-Space Search}
Theories of creative cognition and scientific reasoning characterize discovery as an iterative movement between two distinct spaces: the \textit{already known} and the \textit{not-yet-known}. For instance, Concept-Knowledge (C-K) theory models design as reciprocal transitions between validated knowledge and unvalidated concepts~\cite{hatchuel2003new}. Similarly, \citet{klahr1988dual} describe scientific discovery as a dual search across interacting hypothesis and experiment spaces, while the Geneplore model emphasizes cycles of generative and exploratory processes for vetting partially developed creative ideas~\cite{finke1996creative}.

In the context of research ideation, this dual-space dynamic takes the shape of a continuous dialogue: a developing idea (the not-yet-known) is shaped by existing literature (the already known), and as the idea evolves, it shifts which literature becomes relevant. \system{} operationalizes this recursive process. By facilitating \textit{literature-initiated pivots}, the conceptual movements between these two spaces, \system{} helps researchers iterate on nascent ideas. It supports this mechanism by segmenting ideas, dynamically clustering relevant literature to each segment, reasoning over the literature space, and incorporating insights back into the idea.

\subsection{Tools for Literature Understanding}

An important part of research ideation is developing a deep understanding of the landscape of related literature. Yet the volume of scientific literature is so large~\cite{altbach2018too} that developing this understanding is often prohibitively difficult. To help researchers read literature at scale, prior HCI systems support activities such as paper comparison~\cite{gu2025abstractexplorer, dang2025corpusstudio, chan2018solvent} and literature discovery~\cite{kang2023synergi, lee2024paperweaver, kang2022threddy, palani2023relatedly, singh2025ai2, kang2022augmenting}. Many of these systems help users schematize a literature space, for example through semantic alignment~\cite{gu2025abstractexplorer}, similarities between papers~\cite{lee2024paperweaver, kang2023synergi}, or methodological similarities~\cite{chan2018solvent}. 

These works demonstrate the utility of organizing a large literature space. The reported effects include lower cognitive load~\cite{gu2025abstractexplorer}, a better understanding of paper relevance~\cite{lee2024paperweaver, kang2023synergi}, and the creation of artifacts that accurately reflect the papers read~\cite{palani2023relatedly}. However, they stop short of carrying those insights into the active development of a research idea. This gap is reflected in the future directions these works identify: ~\citet{palani2023relatedly} note that real-world information seeking often involves multiple queries and ill-defined goals, conditions that resemble research ideation, and suggest extending support for such tasks; and ~\citet{kang2022augmenting} note future opportunities to study literature search in the context of ideation outcomes. \system does this by dynamically clustering and ranking literature based on the specific part of the idea a researcher is revising. For example, if a researcher is refining their proposed contribution, \system{} anchors its reasoning on that specific facet, surfacing and comparing the contributions of adjacent papers so the researcher can articulate their idea's novelty.

\subsection{\hita{Tools for Literature-Supported Ideation}}

A growing body of HCI work explores scaffolding research ideation with LLMs, contributing frameworks for human-AI co-creation of research questions~\cite{liu2024ai} and accounts of AI-supported ideation and creativity support~\cite{shaer2024ai, suh2024luminate, liao2023designerly}. These systems focus on idea generation without necessarily grounding it in the knowledge on which it builds. In research ideation, where an idea's merit is defined by its relationship to prior literature, this grounding is critical.

This is not to say that the concept of ``changing an idea'' has been unexplored. A complementary set of tools actively supports idea refinement and assessment. Closest to our work is Scideator~\cite{radensky2024scideator}, which lets users input a set of research papers, decomposes them into facets (purposes, mechanisms, evaluations), allows users to mix-and-match these facets to form new ideas, and then evaluates their novelty. Similarly, IdeaSynth~\cite{pu2025ideasynth} maps idea facets to nodes on a canvas, enabling researchers to explore combinations through AI-based question-answering over user-provided literature. There are a number of ways that \system differs from these tools. Most importantly, \system focuses on the reciprocal relationship between understanding a literature space and developing an idea: the core mechanism of the literature-initiated pivot. This dynamic is largely elided in IdeaSynth and Scideator, where the main contributions lie in generating many variations of an idea. 
While the literature might be consulted in this generation, dynamically exploring and reacting to it is not the focus of these interactions. This difference is evident in the documented limitations of both systems. Scideator~\cite{radensky2024scideator} retrieved a small number (10) of papers per idea and reported that users stated that the system classified ideas as novel too often and missed important related works. IdeaSynth~\cite{pu2025ideasynth} reports that literature curation in the system is largely manual. Participants felt its recommendations were too limited, making other work hard to discover. 
The outcomes evaluated with these systems reflect these differing goals. IdeaSynth states that users explored more alternative ideas using the system but did not observe any significant difference in users' reported confidence in their understanding of the literature, and Scideator reports increased creativity support (and did not measure difference in literature understanding). \hita{We evaluate \system on the outcome its mechanism targets: participants reported significantly higher understanding of the literature, and experts rated their ideas as significantly more grounded in relevant literature. We consider groundedness as one signal for evaluating early-stage research ideas.}


\section{Understanding literature-initiated pivots}
\label{sec:formative}

We conducted an analysis of research meeting notes and a formative study to gain insight into the following research questions:

\begin{itemize} 
    \item FQ1: When and how do literature-initiated pivots occur during the process of developing a research idea?
    \item FQ2: What challenges do researchers encounter when making literature-initiated pivots?
\end{itemize}

\subsection{Understanding when and how literature-initiated pivots occur (FQ1)}
\label{thematic-formative}
We conducted a document analysis of meeting notes (obtained with permission) from four completed computer science research projects (two HCI and two NLP) at the authors' institution. The first author did an iterative open coding on these notes~\cite[chapter 5]{ref:blandford2016qualitative}, noting when literature-initiated pivots occurred~\footnote{We identified a literature-initiated pivot as any instance where the meeting notes documented a researcher citing a specific paper as the rationale for changing a facet of their idea, or where the notes indicated that a particular paper or literature space should be reviewed to inform a change to the idea. The de-identified coding summary will be made available in supporting materials.}.

Meeting notes referenced literature in an average of 7.5 meetings per project ($SD=2.65$). 
Over the four projects, we identified 21 literature-initiated pivots. We observed two main types of pivots: pivots where the research idea differentiates itself from existing literature ($n=9$) to articulate novelty, and pivots where the idea aligns with literature ($n=12$) to adopt best practices or strengthen validity. We go into detail about literature-initiated pivots that occurred early on in the projects, as these were the situations that \system was designed to support.

\subsubsection{Differentiating from existing literature} In these pivots, researchers refined their idea to establish its novelty relative to literature that closely mirrored their proposed contribution. For example, one project initially referenced literature about identifying relations between individual claims, such as detecting a contradiction between ``The dog is blue'' and ``The dog is red.'' Later, an author shared a paper suggesting that this problem was largely solved, but pointing to an open challenge in resolving information across hierarchies, such as relating high-level summaries to granular sub-claims. The project then pivoted from identifying claim relations to developing claim hierarchies. This illustrates how the idea was revised to differentiate it from existing literature.

\subsubsection{Aligning with existing literature} Authors also referenced existing literature that was similar to parts of their idea. For example, the authors of one project were developing a system that relied on AI for specific text and image extraction tasks. The authors referenced a canonical paper's evaluation as being one to adopt. In this case, the authors sought out literature that was similar in one dimension, evaluation methods, to the project.

\subsection{Understanding challenges in making literature-initiated pivots (FQ2)}

We recruited 5 researchers (3 HCI, 2 NLP; 3F, 2M; median age = 27) via our organization's Slack channels and professional connections for a formative study. These researchers had a median of five years of research experience and were actively engaged in developing a research project. 
Each participant brought a research idea in progress, but not yet fully developed, to the study.

During the session, participants developed a short research proposal across three phases. In Phase 1 (10 minutes), participants used only a document editor to articulate their initial ideas. In Phase 2 (15 minutes), participants refined their ideas using standard search tools. This allowed us to observe how they naturally integrate literature into their ideation without augmentation. In Phase 3 (15 minutes), participants used a design probe~\footnote{Built off of~\cite{vasu2025hyper}, the design probe was an LLM-based system that analyzed the quality of the research idea. It generated an initial evaluation, recommending changes based on referenced literature, and producing arguments for and against the idea's core claims (see Appendix \ref{design-probe}).} that provided literature-grounded feedback on the idea. This phase evaluated the benefits and challenges of using literature-informed support to revise ideas.

All sessions were conducted over Google Meet by the first author. The first author analyzed the transcripts using open coding followed by thematic analysis~\cite[Chapter 5]{ref:blandford2016qualitative} to identify common challenges and opportunities. The analysis was refined through group discussions with the other authors.

Two participants pivoted their idea during Phase 2. All five participants pivoted their idea during Phase 3. This suggests that literature-initiated pivots happen even within a relatively short duration (approximately 30 minutes across Phases 2 and 3). We detail the results of our formative study below. We identified two primary challenges: balancing the tension between expanding on the idea and critically evaluating it against related literature, and pivoting when literature challenges the novelty of the idea.

\subsubsection{CH1: Tension between developing the idea and consulting related literature}
\label{subsec:ch1}

When searching and reading literature in Phase 2, participants had two goals: curating the foundational work their idea builds upon ($n=3$), and showing how their idea differs from it ($n=5$). Both proved challenging. Consider the case of P4. After drafting an idea, P4 wanted to check if existing work had already proposed a similar contribution. Finding three such papers, they read each to note its motivation and contribution, then copied over the papers' future work and limitations sections. Reviewing these notes prompted P4 to revise their idea's methods and evaluation. By the end of the task, they reflected, \emph{``before, I felt very strongly about this project...but this has made me question the novelty of it.''} This suggests a need for support when researchers are concurrently developing and evaluating their ideas against prior work.

\subsubsection{CH2: Pivoting when literature challenges the idea's novelty}
\label{subsec:ch2}

A central part of evaluating an idea is articulating its novelty, especially when existing literature raises doubts about it. Participants described the kind of support they wanted in these moments. Consider the case of P3. After reading two papers that substantially overlapped with their core contribution, P3 found themselves in what they called a ``research rut.'' To get out of it, they wanted guidance on \emph{``how to recommend concretely what I can do to fix this problem in a novel way.''} Importantly, P3 did not want a completely new idea; instead, they said they \emph{``want [support] to prompt me to think about certain aspects of the research problem, rather than thinking for me.''} This suggests that support should go beyond surfacing relevant literature, and instead provide specific, literature-backed ways to articulate an idea's novelty over prior work.

\subsubsection{Additional challenges}

\hita{Our formative work surfaced other difficulties researchers face when engaging with literature during ideation. One was discovering literature that participants did not know how to search for: P1 described the anxiety of unknown related work, \emph{``the fear is there are demons out there lurking around in the related work\dots like they did exactly this before.''} Another was the gradual internalization of literature, which unfolded over longer than a single session: P5 described ideation as an interleaved process of \emph{``read literature and write, read literature and write,''} until \emph{``there's enough literature sustaining the motivation of the [work].''}}

\hita{We designed \system\ around CH1 and CH2 because they were the most-observed challenges specific to literature-initiated pivots. \system's cited-paper expansion (Section~\ref{sec:rel-evidence}) also partially supports discovery, a complementary challenge to internalization. However, \system was not designed to support literature discovery and internalization beyond the scope of CH1 and CH2.} 

\begin{figure*}[t] 
    \centering
    \includegraphics[width=\linewidth]{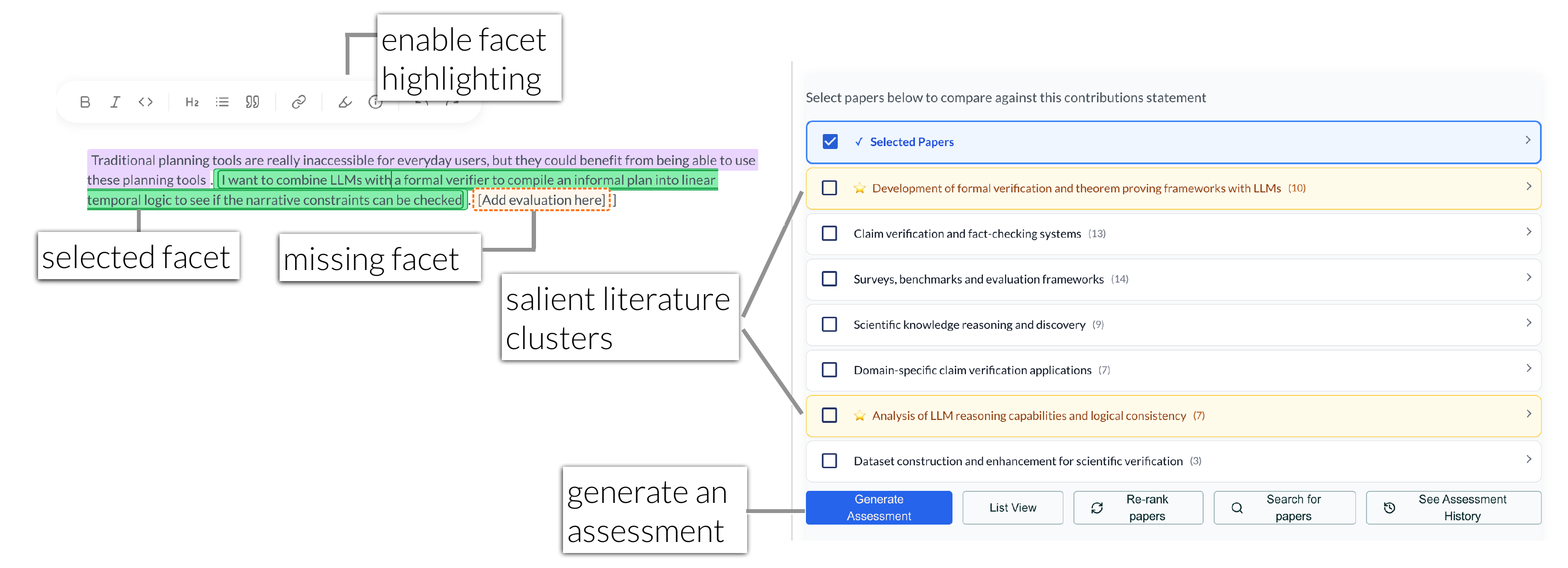}
    \caption{\normalfont \textbf{The \system interface.} The left pane is a document editor with a toolbar toggle to highlight facets. The right pane is a paper panel that surfaces literature clusters. The interface colors text by idea facet: problem (purple), contribution (green), and evaluation (orange; currently missing). Clusters most relevant to the selected facet (here, the contribution statement) are starred. After selecting papers, the researcher can generate an assessment to compare the idea against selected work.}
    \Description{The LitPivot interface with explanatory callouts. Left Pane (Editor): A callout points to a toggle to "enable facet highlighting." The text below is color-coded: a purple sentence (problem), a green sentence marked "selected facet" (contribution), and an orange placeholder marked "missing facet" (evaluation). Right Pane (Literature): Displays a list of related paper topics. A callout for "salient literature clusters" points to two starred topics highly relevant to the selected green facet. At the bottom, a callout points to a blue "generate an assessment" button.}
    \label{fig:system}
\end{figure*}

\subsection{Design goals}
We identify two key opportunities:

\begin{itemize} 

    \item \emph{DG1: Support dynamic curation to align and differentiate an idea.} Our document analysis (Section~\ref{thematic-formative}) showed that researchers seek literature both to align with and differentiate from specific parts of their idea. \system{} supports this by iteratively decomposing the evolving idea into facets and organizing the literature around them (Section~\ref{sec:organizing-literature}), helping researchers see the exact subset relevant to the part of the current idea they want to develop.

    \item \emph{DG2: Use literature-based suggestions to inspire literature-initiated pivots.} Participants reported a tension between developing their ideas and consulting literature to refine them (Section~\ref{subsec:ch1}), especially when prior work threatened their idea's novelty (Section~\ref{subsec:ch2}). Participants wanted more than conflicts surfaced; they wanted concrete recommendations for how to pivot. Therefore, \system{} goes beyond flagging overlaps, applying graph-based checks and lightweight formalizations over the literature space (Section~\ref{sec:graph-based-check}) to suggest specific, actionable rewrites to the idea.
    
\end{itemize}
\section{System}\label{System}



\hita{\system{} can be used to help researchers develop an initial idea in dialogue with the literature.} \system{} features a text editor where researchers can further develop their ideas (Figure~\ref{fig:system}) and break them down into specific facets. Following prior work on faceted research ideation~\cite{pu2025ideasynth, radensky2024scideator}, we structure these as \texttt{problem}, \texttt{contribution}, and \texttt{evaluation} statements. These facets serve as clickable conduits for checking specific parts of the idea against the literature. When a researcher selects a facet, \system{} dynamically clusters the literature space around that specific component, visually highlighting the most relevant clusters. After the researcher selects a subset of these papers, \system{} evaluates the facet against the literature. For example, to evaluate a \texttt{problem} statement, \system{} extracts problem statements from the selected papers to check if the proposed idea addresses a problem that is of interest to the research community. Conversely, to evaluate a \texttt{contribution}, \system identifies how it differentiates itself from existing work (detailed via our bipartite graph approach in Section~\ref{sec:bipartite-graph}). Finally, alongside a report on the facet's literature fit, \system{} provides concrete, actionable suggestions for revising that facet.

\subsection{Narrative scenario}

We describe the experience of interacting with \system in a narrative walkthrough. Cy, an NLP researcher studying LLM evaluation on complex reasoning tasks, wants to develop this idea: using LLMs to translate informal plans into precise specifications that verification tools can test for correctness. She knows the NLP literature well but is less familiar with the relevant formal methods work. She turns to \system and types her initial idea into the text editor.

\paragraph{Exploring faceted literature space}
After writing out her idea, Cy presses a modal key, triggering \system to retrieve all related papers and list them in the right-hand panel. Cy wants to know whether her contribution is novel. She clicks the ``Enable facet highlighting'' button, which divides her idea into facets.

She clicks on the \texttt{contribution} segment of her idea. The side panel reorganizes the literature into labeled clusters of prior contribution types, such as LLM-based formal verification frameworks or claim verification systems. These clusters help Cy understand the contribution types in the existing literature. The clusters most relevant to her proposed contribution are starred (see Figure \ref{fig:system}).



\paragraph{Assessing and updating an idea facet}
Cy selects papers in clusters to assess her proposed contribution against and generates an assessment. \system notes the proposed contribution's overlap with prior work: \textsc{VeriPlan} \cite{lee2025veriplan} ``already couples an LLM, a rule translator, and a model checker.'' This tells Cy that her contribution as written is likely not novel. The assessment also highlights an open limitation articulated in another paper \textsc{LeanReasoner} \cite{jiang2024leanreasoner}: ``the brittleness in commonsense formalization.'' This intrigues Cy. To understand this gap more before pivoting, she reads an evidence snippet from  \textsc{LeanReasoner}'s discussion section. She learns that while prior work verifies rigid task plans, it fails when constraints rely on implicit, real-world knowledge.

She reviews \system's suggested rewrite, which proposes translating free-form creative narratives into formal constraints. Cy reflects on how narratives rely on the exact commonsense reasoning that \textsc{LeanReasoner} struggles with and reads through the \textsc{LeanReasoner} paper to get a better understanding of its methods. Cy reframes her research contribution. Instead of her original idea of translating rigid plans into specifications, she pivots to using an LLM to translate free-form narratives into temporal logic, using a model checker to verify the translation. Because her contribution has shifted in both domain and type, \system dynamically updates the literature space. New clusters of papers become relevant, including work on evaluating LLMs in creative writing tasks and on combining LLMs with temporal logic. This updated view saves Cy from the friction of restarting her manual search from scratch. Instead, she can stay focused on developing her idea while the literature dynamically adapts to her idea.

\paragraph{Maintaining idea congruency} 
Cy must now re-articulate what problem motivates a system like this. \system brings this to her attention by flagging the now-outdated \texttt{problem} facet, signaling that the idea's internal congruency has broken.

Cy clicks the \texttt{problem} segment of her idea. \system{} organizes the literature into clusters based on the problems addressed in prior work. She selects several papers that relate to the problem her new contribution might address, such as inconsistencies in free-form narrative generation, to check her idea against. \system surfaces a well-documented, open challenge in that literature: maintaining plot lines in long narratives often exceeds an LLM's context window. Seeing the direct connection between this documented gap and her new direction, Cy sees that her contribution is well-suited to solve it. She uses insights from \system to rewrite her \texttt{problem} statement, focusing on how formal verification can preserve temporal consistency in long-form storytelling, a task that heavily relies on the commonsense reasoning she identified earlier. Cy now feels ready to further develop this refined idea. Ultimately, \system supported Cy in making several literature-initiated pivots: reshaping an already-existing contribution into one that addresses an open problem in the research community.

\begin{figure}[t] 
    \centering
    \includegraphics[width=\linewidth]{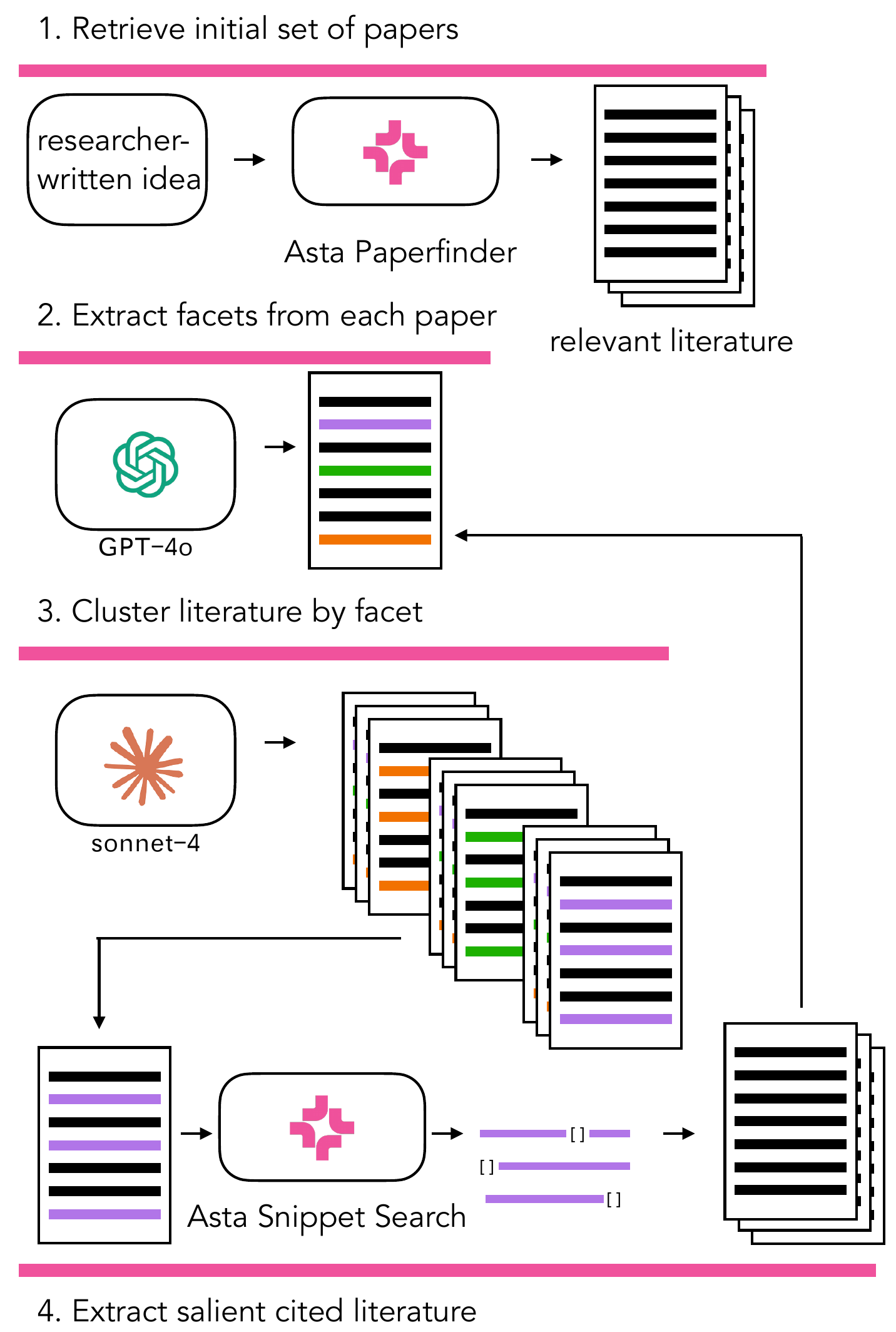}
    \caption{\normalfont \textbf{\system pipeline for literature retrieval and faceted paper clustering.}
    (1) The researcher’s idea is used to retrieve an initial set of relevant papers via Asta PaperFinder. (2) GPT-4o extracts idea-relevant facets from each paper. (3) Sonnet-4 clusters the literature by facet, surfacing groups of papers most relevant to each part of the idea. (4) The system extracts papers that are meaningfully cited within the extracted facets and includes this in the literature set.}
    \Description{A four-step flowchart illustrating the LitPivot pipeline for literature retrieval and clustering: Step 1: Retrieve initial set of papers. A "researcher-written idea" is processed by Asta PaperFinder to output a stack of "relevant literature." Step 2: Extract facets from each paper. GPT-4o analyzes the literature and highlights color-coded facets within the text. Step 3: Cluster literature by facet. Sonnet-4 organizes the papers into distinct stacks grouped by their specific color-coded facets. Step 4: Extract salient cited literature. Asta Snippet Search pulls cited snippets from the faceted text to discover new papers. An arrow loops these new papers back to Step 2, indicating an iterative cycle.
}
    \label{fig:paper_pipeline}
\end{figure}

\subsection{Constructing the literature space} 
\label{sec:paper-space}

The literature space was organized to: 1) curate a large set of relevant literature, and 2) inspire inquiry into the details most useful to the researcher. The pipeline is in Figure~\ref{fig:paper_pipeline}.

\subsubsection{Retrieving relevant literature}
\label{sec:rel-evidence}

\system passes the researcher's idea as a seed query to Asta Paper Finder\footnote{Paper Finder, \url{https://github.com/allenai/asta-paper-finder}}, an open-sourced LLM-powered scientific search engine. Paper Finder retrieves an initial set of relevant papers by Semantic Scholar ID. The IDs are passed into the Semantic Scholar Snippet Text API\footnote{\url{https://api.semanticscholar.org/api-docs/graph\#tag/Snippet-Text}} to retrieve the full text for each paper. Each paper's limitations, future work, contributions, motivations, evaluations, and methods sections are extracted. If an extracted section cites a work not already in the literature set, \system adds the referenced paper to the corpus. The final literature set thus includes both the most directly relevant papers and the foundational work that informed them. An LLM categorizes each paper based on the researcher's idea\footnote{Prompt adapted from Asta Paper Finder: \url{https://paperfinder.allen.ai/}} as perfectly relevant, somewhat relevant, complementary, or not relevant. In the UI, papers appear in descending order of relevance. Researchers can request additional literature at any point, either by adding papers via the Semantic Scholar Academic Graph API\footnote{\url{https://api.semanticscholar.org/api-docs/graph}} or by prompting the system to fetch new results based on a revised idea.

\subsubsection{Organizing the literature space}
\label{sec:organizing-literature}

To surface the literature most useful to the researcher, \system{} clusters papers conditioned on specific facets of the research idea. \system{} groups extracted problem, contribution, and evaluation sections from related work. Papers sharing similar facets are then clustered and labeled by type (Appendix~\ref{ap-cluster-contribution}). To surface the most relevant clusters during editing, \system identifies the clusters most useful relative to the highlighted facet (Appendix \ref{ap-identify-relevant-cluster}). Literature-initiated pivots may shift which papers are most relevant, and researchers can trigger an update to refresh the literature view. When the user clicks the re-rank button in the global view, \system recalculates paper relevance (detailed in Appendices \ref{ap-cluster-contribution} and \ref{ap-identify-relevant-cluster}). We designed this as a user-initiated interaction to manage system latency and give researchers control over when to reorganize the entire literature space around their revised idea.


\subsection{Supporting the literature-initiated pivot}
\label{sec:graph-based-check}

We employ graph-based checks and retrieval-augmented generation to check an idea facet against selected literature.

\subsubsection{Align facet with literature}

Our formative study revealed that participants seek literature to either motivate a problem or adopt an evaluation methodology. Inspired by \cite{ellis2008framework}, we assert that a robust problem statement must: (1) be grounded in citable prior work, (2) establish the problem's existence through evidence of failures, gaps, or pain points, and (3) argue for its significance through impact, risks, or opportunity. A robust evaluation statement must: (1) assess whether the contribution aligns with the problem, (2) test whether it answers the research questions, and (3) verify its plausibility against how similar contributions are evaluated in prior work.

These criteria are translated into prompts (see Appendices \ref{ap-eval-eval} and \ref{ap-problem-eval}). Since \system prioritizes literature-driven framing, the problem assessment checks whether the draft can cite at least one matching statement from the selected papers. The evaluation assessment conditions on the proposed problem, contribution, and few-shot examples of how selected papers evaluate similar work. As prior work may sometimes use suboptimal methods, \system surfaces literature patterns for the researcher to review and selectively adopt. Researchers can accept or reject suggestions and request additional context to steer the search. Following these assessments, an LLM generates literature-informed rewrites: the problem rewrite must specify (1) the problem, (2) evidence it is real, and (3) why it matters; the evaluation rewrite must specify (1) whether the evaluation shows the contribution addresses the problem, and (2) whether the plan is feasible and sensitive enough to detect intended effects.

\subsubsection{Differentiate facet from literature}
\label{sec:bipartite-graph}

A proposed contribution must articulate its relationship to and distinction from existing work, a task our formative study identified as particularly challenging (Section ~\ref{sec:formative}). We formalize this by constructing a bipartite graph: one vertex set represents contributions extracted from selected papers, the other represents their limitations or future work statements, and edges indicate that a contribution addresses a specific limitation. An LLM prompt constructs this graph by taking all extracted contributions and limitations/future-work statements as input and returning a bipartite graph in JSON. A second prompt links the proposed contribution to known limitations and future work: links solely to already-addressed items suggest an incremental contribution, replication, or extension, while links to unaddressed items suggest conceptual novelty (see Appendices \ref{ap-bipartite-graph} and \ref{ap-sol-eval}). \system then rewrites the contribution against three criteria: (1) does it directly address the stated problem, (2) is success plausible given the methods and constraints, and (3) how is it positioned relative to the linked limitations and future work?

\subsubsection{Maintaining consistency between facets}

Because the \texttt{problem}, \texttt{contribution}, and \texttt{evaluation} facets that make up a researcher's idea are interdependent, if a researcher revises one facet, they might need to go back to other parts of the idea and revise them to make sure the idea is internally congruent. When a facet is changed, \system passes the revised idea, the literature analysis, and the newly edited facet to an LLM, instructing it to identify which other facets should be updated to maintain consistency. If a mismatch is detected, for example, if a newly revised contribution no longer addresses the original problem or requires different evaluation metrics, \system visually highlights the affected facets to prompt the researcher for further revision (Appendix \ref{ap-id-affected-facet}). Additionally, after segmenting the initial idea, \system flags any core facets that have not yet been articulated, prompting the researcher to draft the missing facet (Appendix \ref{identify-missing}). A researcher can also request a full assessment of the entire idea at any time (Appendix \ref{ap-gen-complete-assessment}).


\subsection{Implementation}

\system is implemented as a web application. The frontend uses the React framework in JavaScript, while the backend is implemented in Python using Flask. Once retrieved, paper data is stored as JSON files. Different language models are used across the pipeline based on the task. Broadly, OpenAI's \texttt{GPT-4o} (temperature=0.7) is used for low-level tasks: segmenting ideas into facets, extracting facets from paper full-text, identifying relevant clusters, detecting missing facets, and flagging affected segments after an edit. OpenAI's \texttt{o3} (temperature=0.7) is used for tasks involving reasoning: graph instantiation and generating evaluations and suggestions for each facet. Anthropic's \texttt{sonnet-4} (temperature=0.7) is used for tasks involving long contexts, such as clustering papers by facet.\footnote{We assigned models by latency rather than defaulting to the most capable model for every step: \system\ is an interactive system, and reasoning models are considerably slower, so we reserved them for the steps that most benefit from deeper reasoning. Users can swap in larger models as desired.}

\subsection{Technical Evaluation}
\label{sec:tech-eval}

\hita{This section assesses the \system pipeline: does a pivot-supporting pipeline produce better literature-grounded research idea feedback than that of prior systems?} We compare \system's pipeline against another literature-based ideation system, IdeaSynth~\cite{pu2025ideasynth}. Rather than benchmarking against the growing landscape of AI writing tools~\cite{openai2026prism}, into which we conjecture \system's pipeline could eventually be integrated, we focus on assessing it against a research-grade baseline. Two domain experts, recruited through professional networks, rated outputs from both systems across 5 seed ideas drawn from their areas of expertise, each decomposed into problem, contribution, and evaluation facets (15 segments total). The LLM and literature corpus were held constant across conditions, with each idea's source paper withheld. Blinded to condition, experts rated on 1--7 Likert scales: (i) how well the evaluation identified strengths and weaknesses relative to the literature, and (ii) whether incorporating the suggestions would yield a stronger paper.

\system\ outperformed the baseline on both measures. We use the Wilcoxon signed-rank test as our primary inference and report paired $t$-tests as a convergent robustness check, including mean differences, 95\% CIs, and paired effect sizes ($d_z$). For suggestions, both raters independently favored \system\ over baseline with large effects (Rater 1: $M{=}3.40{\to}5.27$, $M_{\text{diff}}{=}+1.87$, $t(14){=}4.00$, $p{=}.001$, $W{=}8$, $p{=}.005$, $d_z{=}1.03$; Rater 2: $M{=}3.73{\to}5.73$, $M_{\text{diff}}{=}+2.00$, $t(14){=}3.94$, $p{=}.002$, $W{=}8$, $p{=}.005$, $d_z{=}1.02$). Aggregating both raters ($n{=}30$ paired judgments) yielded a mean improvement of $+1.93$ points ($t(29){=}5.71$, $p{<}10^{-4}$; $W{=}30$, $p{=}7.1{\times}10^{-5}$; $d_z{=}1.04$). For evaluations, one rater showed a significant preference for \system\ ($M{=}3.53{\to}5.47$, $M_{\text{diff}}{=}+1.93$, $t(14){=}2.91$, $p{=}.012$, $W{=}14.5$, $p{=}.016$, $d_z{=}0.75$), while the other showed a non-significant trend ($M{=}3.60{\to}4.40$, $M_{\text{diff}}{=}+0.80$, $t(14){=}1.52$, $p{=}.15$, $W{=}20.5$, $p{=}.14$, $d_z{=}0.39$); the combined analysis was significant with a moderate effect ($M_{\text{diff}}{=}+1.37$, $t(29){=}3.18$, $p{=}.004$, $W{=}66$, $p{=}.005$, $d_z{=}0.58$). This suggests that our pipeline's design for supporting literature-initiated pivots has the intended positive effects of 1) correctly relating the idea to a selected literature space, and 2) providing actionable ways the idea can be better situated within that space.~\footnote{Inter-rater reliability was low across conditions (Krippendorff's $\alpha$ ranging from $-0.33$ to $0.11$), consistent with prior work on the subjectivity of single-item scholarly judgments~\cite{shah2018design, bornmann2010reliability, pier2018low}. Given this, we emphasize within-rater paired contrasts over absolute scores, and interpret the evaluations as such.}

\section{Study 1: User Study}\label{results}


\paragraph{Participants}
We recruited 17 researchers in computer and information sciences whose research aligned with the study tasks using professional networks and social media. Because \system aims to help researchers concurrently evolve an idea and their understanding of a literature space, we primarily targeted early-career researchers. While these individuals possess foundational research skills, they presumably benefit more from literature scaffolding than senior researchers who already possess rich, internalized domain knowledge~\cite{fitzgerald2017information, davidson2008provenance}. Accordingly, our participants consisted mostly of PhD students (82\%), along with one postdoc and two industry researchers. Participants had varying levels of research experience: 59\% reported 2–5 years and 41\% reported 6–10 years. Their research areas spanned NLP (9), HCI (3), robotics/RL (2), biomedical/health AI (2), and ML/AI (1). Participants also reported varying use of AI tools for ideation: 18\% always, 35\% frequently, 35\% occasionally, and 12\% rarely.

\paragraph{Study setting}
The study was conducted over Google Meet. Sessions lasted 90 minutes, and participants received \$75 USD.~\footnote{The study was approved by an internal review board. Our study protocol and materials are included in the supplemental material}.

\paragraph{Interfaces}
In the baseline condition, participants used a  QA interface that let them select papers and ask questions (see Figure \ref{fig:baseline}). To isolate the effect of literature-initiated pivots, this baseline did not segment the idea into facets or support targeted pivots through automated assessments, unlike \system. Instead, it reflected a common alternative: a standard text editor paired with RAG-based LLM assistance. Its backend matched the literature-grounded QA feature in IdeaSynth~\cite{pu2025ideasynth} and used the same paper corpus as \system. The QA interface answered questions using the full text of the selected papers and the current idea draft as context.\footnote{We counterbalanced condition order to control for learning and ordering effects. Although \system normally supports adding new papers, we did not allow participants to add their own during the study to ensure both conditions used the same paper set.} \hita{Both conditions used the same set of ${\sim}100$ papers, retrieved by Asta PaperFinder~\cite{paperfinder}, an agentic search system with sub-agents for query analysis, planning, and iterative retrieval expansion.} Before each task, participants received a tutorial for the assigned condition. The experimenter demonstrated the interface's features and asked participants to practice using them by briefly editing a sample idea.

\paragraph{Test Tasks}
\label{study_tasks}
Each participant developed two research ideas, one with \system and one with the baseline. Based on participants' reported familiarity with subareas of HCI and NLP, they were assigned two seed ideas~\footnote{An example task topic, ``biomedical claims verification,'' used the following seed idea: ``Verifying biomedical claims is difficult, especially when you have to look through a paper's citations, and those citations' citations, to get to the actual study that conducted the experiment that yields primary evidence in support of a claim. We will create a dataset of biomedical claims and the citation `hops' it takes from paper to paper to entirely verify the claim.''}. We used standardized seed ideas rather than using participants' own ideas to control for differences in project maturity. Participants had 30 minutes per task. We collected:

\begin{enumerate}
    \item \emph{Behavioral interaction logs}. We recorded logs of text edits, papers selected or retrieved, assessments generated (in \system), and questions asked (in the baseline).
    
    \item \emph{Self-reported data}. Participants rated their agreement on 7-point Likert scales (1=strongly disagree, 7=strongly agree) across several dimensions. To measure how perceptions of the ideas changed, we collected pre- and post-task ratings for: (1) perceived novelty, (2) perceived feasibility, and (3) perceived utility of the idea. Additionally, participants provided post-task ratings across three additional metrics: (1) understanding of the literature space, (2) helpfulness of the system's outputs, and (3) trust in those outputs. We corrected families of related outcomes using the Holm--Bonferroni method. We report estimated marginal means with 95\% CIs and effect sizes; for non-parametric data, we use Wilcoxon signed-rank tests with Cliff’s $\delta$.
    
    \item \emph{Qualitative data}. At the end of the study, participants reflected on their experiences during both tasks. The lead author analyzed participants' responses through a thematic analysis process~\cite[chapter 5]{ref:blandford2016qualitative}. Analyses were refined through discussion with another author to derive key themes.
    
    \item \emph{Expert evaluation of ideas}. The last two authors, who are domain experts in the areas the ideas were from, rated the final ideas, blind to condition. Assessing research ideas requires long-term execution to judge feasibility or impact~\cite{pu2025ideasynth,boudreau2016looking, wang2013quantifying}, so we focused on one criterion that encapsulates the primary effect of \system: ``\emph{This idea is well-grounded and well-argued using relevant literature}.'' Ratings used a 7-point scale. We measured agreement with Krippendorff's alpha and iteratively refined the guidelines to reach consensus. Agreement improved from 0.389 to 0.804 after three rounds of discussion and guideline refinement (about 140 minutes total; Appendix~\ref{apx:guidelines}); remaining conflicts were resolved.

\end{enumerate}

\hita{We scope our measures of idea novelty, feasibility, and utility to self-report data. Judging the significance of a research idea before it has been executed can be noisy~\cite{pier2018low}; self-report data can be assessed at the idea stage and follows practices in prior ideation work~\cite{pu2025ideasynth, si2025can}. For an early-stage idea, whether the researcher perceives it as novel, feasible, and useful itself might be a practical and useful signal. Measuring downstream idea significance and impact is an important direction for future work (Section~\ref{future-work}).}


We hypothesized that:
\begin{itemize}
    \item H1: \system helps researchers understand the literature space better than the baseline.
    \item H2: \system helps researchers generate better-grounded ideas than the baseline.
\end{itemize}

\subsection{\system{} helps researchers better understand the literature space (H1).}

Participants selected significantly more unique papers with \system{} to generate assessments (median=7, $M=7.1, SD=3.9$) than they did for Q\&A in the baseline (median=3, $M=4.4, SD=4.3$; $W=15.0, p=0.010$). While an increased document selection count might not indicate quality of interaction, our proxies showed some evidence that participants better understood the literature space that they engaged with. Participants reported a significantly better understanding of the literature space with \system ($M=5.71$) than with the baseline ($M=3.88$; $t(16)=5.85, p<.001, d_z=1.42$).

\paragraph{Literature search in the baseline}
In the baseline, participants ($n=13$) typically chose the first few papers in the interface ($M=2.92$ papers, $SD=1.19$) to ground the chat interaction. A smaller number of participants ($n=4$) read through the list of papers first and selected a larger set of papers ($M=9.25$ papers, $SD=7.23$). On average, participants asked 3.9 questions in the baseline. Common question types were requests to evaluate the idea (13; ``Is this idea novel?''), summarize the literature (10; ``What do these papers do?''), and clarify details (7; ``What benchmarks do the paper report?'').

\paragraph{Selective literature pruning} 
Using \system, we observed 11 participants follow an iterative literature selection process: beginning with all papers in starred clusters, then narrowing the set by deselecting papers, and in some cases adding papers from non-starred clusters ($n=6$), for subsequent assessments of the same idea facet. Participants revised their idea based on gaps identified in the literature. For example, P1 was editing a contribution about generating ``synthetic medical notes that are consistent and factual,'' initially selecting 12 papers across three starred clusters. After \system identified ``temporally coherent multi-visit corpora'' as an open problem, P1 reviewed abstracts, deselected six papers focused on single-visit notes, and revised the contribution. After another assessment, P1 further refined the contribution to a technique for ``check/reward signals for contradictions across multiple hospital visit notes.'' By restructuring their literature space and acting on specific gaps surfaced by \system, P1 pivoted their contribution.

\paragraph{Expanding literature assessment}
Six participants increased the number of papers used across assessments as they refined their ideas. For example, P9 generated four assessments using 2, 3, 4, and finally 6 papers. Their final idea reflected this expanding literature set: it used statistics from prior work to motivate the problem, combined technical contributions from two papers into a novel contribution, and drew on another paper to ground the evaluation.



\begin{figure}[t] 
    \centering
    \includegraphics[width=\linewidth]{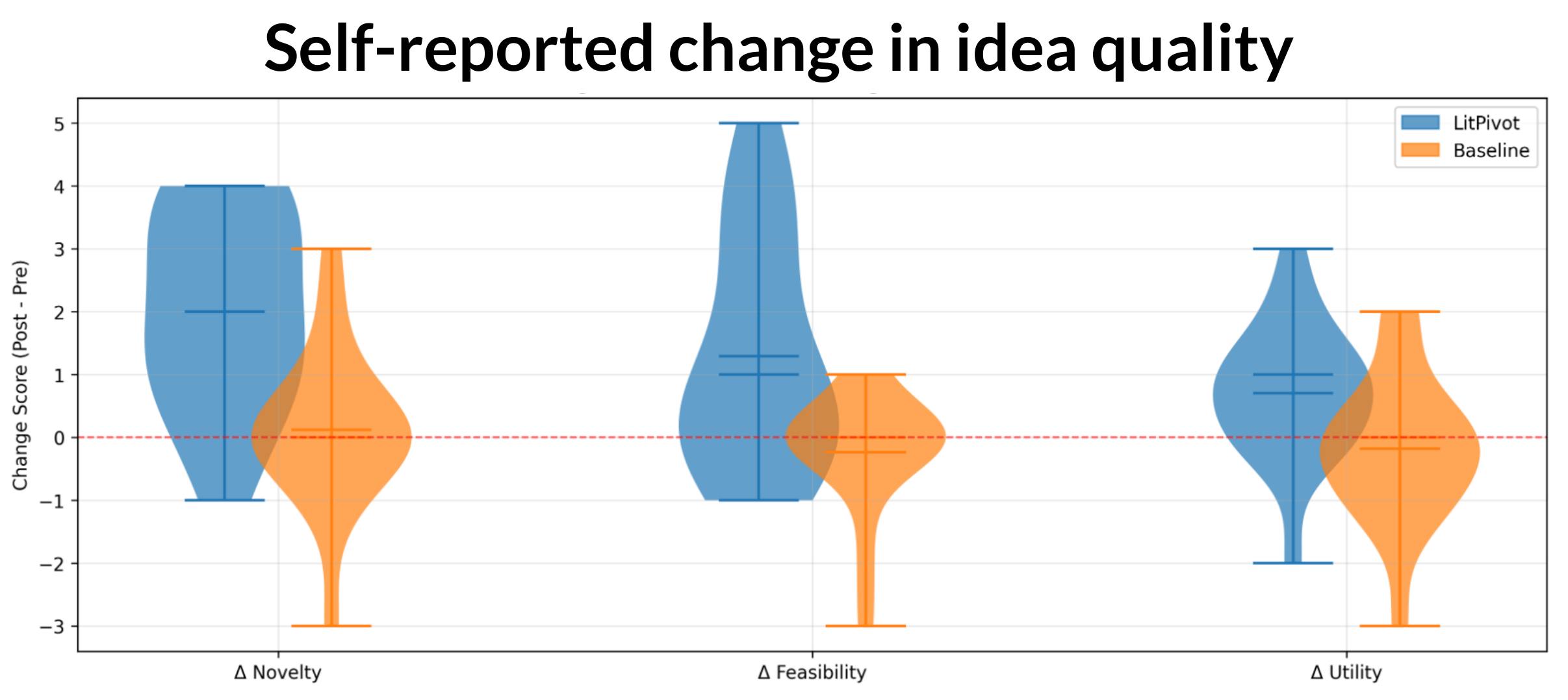}
    \caption{\normalfont \textbf{Distribution of change in Likert scores} (post task - pre task) for perceived idea novelty, feasibility, and utility across conditions. Horizontal bars indicate median and quartile values; the red dashed line marks no change.}
    \Description{A violin plot compares self-reported change in idea quality between two conditions: LitPivot (blue) and Baseline (orange). Three metrics are shown on the x-axis: change in Novelty, Feasibility, and Utility (post–pre scores). For all three metrics, LitPivot distributions are wider and centered above zero, indicating greater positive change, whereas Baseline distributions are narrower and closer to or below zero, suggesting little to no improvement. Horizontal lines within each violin indicate the median and interquartile range, and a red dashed line at zero marks no change.}
    \label{fig:results}
\end{figure}

\subsection{\system helps researchers generate better-grounded ideas (H2).}


\paragraph{\hita{Improvements in self-reported idea qualities}}
\label{sec:findings-quality}
Analyzed as per-participant change scores, participants reported that their ideas improved in the following dimensions (see Figure \ref{fig:results}):

\begin{itemize}
    \item Participants reported a significantly larger increase in their idea's novelty with \system\ ($\Delta=2.29$ vs.\ $\Delta=0.47$; $t(16)=3.23$, $p_{\text{corr}}=.016$, $d_z=0.78$), with a mean difference of 1.82 (95\%\ CI $[0.63,\,3.01]$).
    \item Participants reported a significantly larger increase in their idea's utility with \system\ compared to the baseline ($\Delta=0.53$ vs.\ $\Delta=-0.76$; $t(16)=2.81$, $p_{\text{corr}}=.037$, $d_z=0.68$), with a mean difference of 1.29 (95\%\ CI $[0.32,\,2.26]$).
    \item Participants reported a larger increase in perceived feasibility, though the difference was not significant ($\Delta=0.88$ vs.\ $\Delta=-0.35$; $t(16)=2.11$, $p_{\text{corr}}=.152$, $d_z=0.51$), with a mean difference of 1.23 (95\%\ CI $[-0.01,\,2.47]$)\footnote{We hypothesize that feasibility was unchanged because participants were assigned ideas in this study and, though they were assigned ideas based on their research areas, they might not have been able to effectively assess feasibility.}.
\end{itemize}

\paragraph{Effect of \system on \hita{groundedness of final ideas}}

To complement self-reported metrics~\footnote{Self-reported metrics are common in HCI ideation research~\cite{pu2025ideasynth, gero2022sparks, suh2023sensecape}, following established Creativity Support Tool evaluation practices~\cite{cherry2014quantifying, carroll2012triangulating}.}, we conducted a blinded expert evaluation of idea grounding. Ideas produced with \system were rated as significantly more well-argued and grounded than those from the baseline. The median rating rose from 2 to 5 out of 7, and the mean doubled (Baseline $M{=}2.59$, $SD{=}1.33$; \system\ $M{=}5.18$, $SD{=}1.85$; both $n{=}17$). A Wilcoxon signed-rank test confirmed higher ratings for \system\ ($W{=}4.5$, $p{<}.001$), and a paired $t$-test showed the same pattern ($t(16){=}5.33$, $p{<}.001$; mean difference $=2.59$, 95\% CI $[1.56,\,3.62]$; $d_z{=}1.29$). See Table~\ref{tab:baseline-vs-litpivot} for examples.

\paragraph{Using \system to make literature-initiated pivots}
\hita{We observed participants using \system to develop their idea facets in two ways. Most participants ($n{=}13$) moved from one out-of-sync facet to another, not necessarily sequentially. For example, P4 first modified a problem statement, the first sentence of the idea. \system flagged that the proposed evaluation, the last sentence of the idea, no longer matched the new problem statement. P4 edited their contribution last after it was flagged as out-of-sync. Alternatively, some participants edited their ideas sequentially, revising one facet multiple times before editing the next ($n{=}4$). For example, P5 generated three assessments for each facet before moving to the next facet segment. These two revision behaviors suggest that \system supports different literature-grounded idea development workflows.}

Participants rated \system's suggestions as significantly higher than the outputs from the baseline in terms of helpfulness ($M=6.24$ vs.\ $M=4.12$; $t(16)=3.96$, $p=.001$, $d_z=0.96$) and trust ($M=6.00$ vs.\ $M=3.88$; $t(16)=5.47$, $p<.001$, $d_z=1.33$). There is some evidence that participants actively iterated on the suggestions rather than accepting \system's outputs as is: across all sessions, $50\%$ of the assessments were incorporated as written, $45.6\%$ were partially incorporated with edits, and $4.4\%$ were not incorporated.

\section{Study 2: Open-ended qualitative study}
\label{sec:qual-study}
To observe how researchers use \system to develop their own ideas, we conducted a qualitative study. Five researchers who had participated in the user study in Section \ref{results} were recruited. Before the session, participants shared a current idea and any relevant literature they had, which we pre-populated into the system. During a 30-minute session, they used \system to develop the idea.

\paragraph{\system was used across the research cycle}

Researchers used \system for vetting the validity of new ideas against relevant literature ($n=2$) and framing mature ideas ($n=3$). Most participants ($n=4$) stated they would reuse the system for research writing. As P12 stated, \textit{``this is super useful...in the early stages of when I don’t know about a research idea and my idea is not fully shaped...and when I do have a fully formed research idea and I want to...make sure I’m not copying someone’s existing research idea.''}

\paragraph{\system helps researchers relate their ideas to literature}
All participants used facet-based clusters to search for literature. P3 stated, \emph{``it's way easier to navigate to a potential[ly relevant paper] when everything's grouped by topics.''} Notably, participants selected different literature when iterating on different facets, highlighting the utility of dynamic literature curation. P2, for instance, initially believed their idea contributed a result specific to fine-tuning. However, after identifying a contribution cluster discussing reinforcement learning (RL) approaches, they reframed their contribution to generalize across both fine-tuning and RL. P2 reflected: \emph{``[\system{}] could be useful for finding similarities between papers\dots especially in this case where this concept from one area\dots has parallels in another.''}

Participants ($n=4$) used assessments to evaluate their ideas' potential contributions, leading to pivots in how they framed their contribution. For example, P3 began with an early-stage idea focused on question generation in a specific high-stakes domain. After generating three assessments, each with a different set of literature, they noted that this initial contribution was too incremental. P3 then pivoted to defining and evaluating ``situational appropriateness'' of question-generation across multiple domains. By the end of the session, they articulated this new contribution, a taxonomy of QA constraints across domains. During the session, they saved seven papers for future reference, incorporated five \system-suggested rewrites to their idea, and copied two evidence excerpts included in \system's output to the idea editor.
\section{Discussion}



\paragraph{How \system's features supported idea development}
Our findings suggest that two design choices of \system contributed to its observed effects. The first is dynamic literature curation. Participants selected significantly more unique papers with \system\ (median 7 vs.\ 3), and the user study revealed two curation behaviors this enabled: selective pruning ($n{=}11$), where participants narrowed starred clusters in response to gaps the system surfaced (e.g., P1's pivot to multi-visit medical notes), and expansion ($n{=}6$), where participants grew their literature set across successive assessments (e.g., P9's four assessments over 2--6 papers). 

\hita{The second is that \system's critiques are specific enough to act on while leaving the decision with the researcher. \system articulates how a facet relates to specific selected papers and proposes a revision, and participants rated these outputs as significantly more helpful ($M$ = 6.24 vs. 4.12) and more trustworthy ($M$=6.00 vs. 3.88) than the baseline's. Researchers also engaged with these assessments: 50\% of assessments were incorporated as written, 45.6\% were edited before incorporation, and 4.4\% were rejected. This pattern of active mediation, rather than passive acceptance, echoes insights from our formative study (P3: wanted support that ``prompt[s] me to think\dots rather than thinking for me'').}

\paragraph{Principles for broader applications}
Existing tools for knowledge-intensive work like clinical reasoning or legal argumentation assume the corpus is static: it can be queried, but it does not change. This assumption shapes more than retrieval interfaces. Search systems rank results by relevance to a query, not by their relationship to the argument a user is developing. Our results suggest a different model, in which the corpus reorganizes as the idea changes and the idea is revised in response to what the corpus surfaces. This model carries two design implications. First, retrieval could be idea-sensitive rather than query-sensitive, modeling the current state of the user's developing argument. Second, feedback could be relational: tools could articulate how a retrieved document bears on the specific claim the user is making. \hita{The model could also extend beyond ideation. Expository writing, as defined by~\citet{shen2023beyond}, produces artifacts that both summarize existing knowledge and generate new insight: authors must understand source texts while revising the artifact they produce, which requires pivots that recalibrate the evidence space and re-contextualize the artifact. Tools could support evidence-initiated pivots broadly, and their effects in these high-stakes settings are a promising future direction.}

\hita{Adapting \system's mechanisms to other fields may require some adaptation. The facet vocabulary (problem, contribution, evaluation) can be used to assess and develop empirical HCI and NLP research. It might need restructuring for fields with different argument structures. The bipartite graph novelty check assumes a literature that explicitly states limitations and future work, a convention that might vary across disciplines and paper types.}

\paragraph{Ethical considerations}
\hita{\system could encourage a particular kind of development of a research idea. For example, the facet structure asserts that an idea consists of a problem, contribution, and evaluation; the bipartite graph asserts that novelty comes from addressing an unaddressed limitation; and LLM-based extraction and clustering might encode further assumptions about the most influential literature relating to the idea. On one hand, these mechanisms set a lower bound on novelty: when a researcher's articulated contribution is too close to prior work, \system\ pushes the researcher to differentiate it, and it does not reject ideas for being unconventional. However, the same grounding that strengthens an idea could result in it being closer to existing literature than the researcher intends, or crowd out early, unformed directions. Paradigm-shifting ideas, in particular, may not correspond to any limitation stated in prior work. Participants in the study edited or rejected half of \system's suggestions rather than accepting them without any edits. Editing \system-generated suggestions depends on researcher diligence. Designers of similar systems should make these embedded values legible to users, and future designs could offer modes with varying degrees of literature-boundedness suited to different phases of ideation (Section~\ref{future-work}).}



\subsection{Limitations}
\label{sec:limitations}

Our evaluation focused on HCI and NLP researchers performing solo, time-boxed tasks largely at a specific stage of the ideation process. While our formative study (Section~\ref{sec:formative}) indicates that ideas can evolve within short time frames, the generalizability of our findings could be limited. \system's utility is contingent on the availability of a robust and accessible literature corpus; its effect may be less apparent when this is not the case. 

\hita{\system{} assesses two dimensions of an idea: how it is grounded in prior work and how it is differentiated from it. Other dimensions (e.g. the significance of a problem, an idea's interestingness, the feasibility of executing it) are left to the researcher, and our evaluation does not measure them. Our formalization of idea differentiation is also one operationalization among many: the bipartite graph treats novelty as coverage of limitations that prior work has left unaddressed, which does not capture novelty that arises from reframing a problem or from work whose limitations are never articulated. \system{} encodes specific values about what makes an idea good, such as grounding and structural coherence, that may not hold across disciplines. Because \system{} does not execute code or run experiments, claims about idea feasibility cannot be verified, and the longitudinal effects of the system on long-term research development remain to be studied.}

\subsection{Future Work}
\label{future-work}

Future work could explore the effects of different literature curation strategies on idea development. \system lets researchers control which literature they evaluate against, in contrast to systems like Scideator that use automated methods~\cite{radensky2024scideator}. Automated literature curation can surface long-tail papers a human might miss, but manual workflows might result in deeper understanding of the literature. Our literature clustering aims to bridge this gap, pointing users to different parts of the literature (framing, method, evaluation, datasets). Future work could further support human-AI collaboration through mixed-initiative search, clustering, and vetting~\cite{feng2024cocoa}. \hita{Additionally, as richer agentic research workflows mature~\cite{feng2024cocoa}, comparing \system against them, and layering pivot support onto them, could be a useful next step.} Longitudinal studies could also track downstream outcomes such as significance and impact, and future designs could let researchers adjust the degree of literature-boundedness across phases of ideation.

Finally, dynamically re-situating evolving text against its sources could extend to other research tasks. For example, a system could provide dynamic feedback that surfaces related research threads to what the user is currently writing. Similar ideas could potentially be applied to update deep research agent reports. This includes tracing citations of existing references to check for new relevant literature published since the report was generated, situating the report within the new documents to generate updated suggestions.
\section{Conclusion}

We present \system, an ideation system that helps researchers iteratively develop a research idea and explore relevant literature. \system supports this process through \textit{literature-initiated pivots}: moments when engaging with relevant literature prompts a researcher to revise the idea's framing, and where that revision changes which literature is relevant. This coupling is driven by reorganizing the literature based on changes to the idea and surfacing whether an idea follows from, departs from, or is already addressed by the  literature. 
Our studies show that ideas produced with \system\ were rated by experts as better grounded in the literature, were perceived by researchers as more novel and useful, and that researchers developed a stronger understanding of the literature space. We argue this points to a broader design principle: a knowledge corpus could be an active participant in shaping new ideas.


\begin{acks}


We thank Dan Weld and Doug Downey for their input on the project framing. Hita thanks the Ai2 intern cohort, especially Yapei Chang, Nishant Balepur, Federica Bologna, Xinran Zhao, Peiling Jiang, and Akhila Yerukola, for making the summer in Seattle memorable.
\end{acks}

\bibliographystyle{ACM-Reference-Format}
\bibliography{refs,andrew-base,andrew-extras}

\appendix

\section{Guidelines for rating idea artifacts}

\label{apx:guidelines}

\begin{itemize}
    \item Focus on whether the idea description leveraged prior work to argue for its novelty and explain its relation to relevant literature. For example, arguments and methods are stronger if they are backed by relevant prior work to show importance or promise.
    \item The idea description is well reasoned (i.e., no jump of logic); Descriptions of problem, contribution, and evaluation should be relevant and coherent.
    \item Do not consider personal research preferences or the ``interestingness'' of the idea.
    \item Assume participants described prior work referenced in their idea descriptions accurately.
    \item Execution feasibility is not part of the evaluation, as it may be highly dependent on the amount of resources the participants have access to.
    \item Rate the ideas holistically, but -1 if the idea description is missing evaluation. (All ideas have background, problem, and contribution statements, but a few were missing evaluation.)
    \item Two example ideas that were agreed upon received a rating of 7 and 1, respectively.
\end{itemize}

\section{Prompts}
\label{appendix:prompts}

\subsubsection{Prompt to segment idea}
\label{ap-segment-idea}

{\ttfamily\small\obeylines A researcher wants help in refining their research idea into a research proposal. To do this, you are tasked with figuring out what parts of their current research idea relates to the following important facets of a research proposal: problem, contribution, and evaluation.
The research problem is any text that describes the core problem a researcher is trying to solve. It answers the following questions: 1. What is the problem this researcher is trying to solve? 2. Why bother solving this problem?
The research contribution is any text that describes the proposed solution to this problem. It answers the following questions: 1. How do current solutions fail at solving the problem? 2. What would a solution to this problem look like? This is the core contribution of the researcher, the core questions the researcher wants to explore in the project. The core contribution might be a proposed system or it might be a study or benchmark that the researchers want to run, or a dataset they are introducing. It is the proposed way that a researcher will answer their research questions. It is what they are developing in this project.
The research evaluation is any text that describes how a researcher will evaluate their contribution. It answers the following questions: 1. How would I know I've solved this problem? 2. How do I plan on solving this problem?
IMPORTANT: The researcher might not have each of these facets, so only classify a section of the text as a particular category if it is the case, and leave the other categories blank. 
IMPORTANT: One segment of text should only have one category.
IMPORTANT: Each segment of text should be a complete phrase or sentence.
IMPORTANT: There should be no overlap between faceted segments.
IMPORTANT: Each segment should be matched and returned verbatim as the input, matching case.
IMPORTANT: If any text states Add evaluation here it should be classified as evaluation.
IMPORTANT: If any text states Add problem here it should be classified as problem.
IMPORTANT: If any text states Add contribution here it should be classified as contribution.
Here is the researcher's idea: \{IDEA\}\par}

\medskip

\subsubsection{Prompt to extract the contribution facet from a paper}
\label{ap-extract-contr}

{\ttfamily\small\obeylines You are helping a researcher understand the core contributions in the provided paper. Extract and return exact quoted excerpts from this paper that represent the core contributions and novelty of this paper. The return text will be a list of exact quotes from the paper. Limit to 3 excerpts. Each excerpt should be 1-2 sentences long.
This is the paper: \{TEXT\}\par}

Note: We used a similar template for extracting other facets (problems, limitations, evaluations, methods, results, future work).

\medskip

\subsubsection{Prompt to evaluate a problem facet}
\label{ap-problem-eval}

{\ttfamily\small\obeylines You will be executing two tasks: first, you will evaluate the researcher's problem statement of their idea and second, you will provide literature-grounded, concrete suggestions for how they can improve their problem statement.
This is their entire idea: \{IDEA\}
This is the part they need to refine: \{IDEA\_SEGMENT\}
IMPORTANT: This is additional context that the researcher wants you to know: \{ADDITIONAL\_CONTEXT\}. If there is no context, ignore this.

To help the researcher refine this part of their idea, here is how current, successful papers describe the problem statements that they are aiming to solve:

\{CONTEXT\}

For your evaluation, you want to make sure that the researcher is answering the following broad questions in their problem statement: 1. What is the problem the researcher is trying to solve? and 2. Why bother solving this problem?

More specifically, with the current formulation of the researcher's problem, can they cite any of the papers above to support their problem statement? Using the excerpts from the papers provided as examples, evaluate whether the problem statement specific enough and if it is an actual problem, and if there is enough support to know that this is the right problem formulation for the researcher to work on. Your evaluation should be no more than 4 sentences, and should include excerpts from and references to as many related papers as possible provide the best assessment of the paper. The references of papers should include the first few words of the paper title and should be of the following form: ```<Paper corpusId="268248445">PaperWeaver</Paper>```.

After the evaluation, you will provide 1-3 concrete suggestions to the researcher on how they can improve their problem formulation, if applicable. The suggested rewrite should be something that the researcher can directly replace their current idea segment with. More formally, rewrite the researcher's idea so that the questions below are clearly answered. Then, include excerpts from relevant literature that could be used to improve the problem formulation. These excerpts should be direct quotes of related problem statements or supporting evidence. If at any point, any of the following questions are currently inadequately answered by the researcher, then you should use the literature provided above and your knowledge of the researcher's idea to ensure that your suggested rewrite answers all of these questions. Here are the questions:

1. What is the problem? This should be of the form "We currently can't do Y" or "We want to do Y"
2. What is the specific problem we are trying to solve?
3. How do we know this is an actual problem?
4. Why should we bother solving this problem?

IMPORTANT: Here is an example of the output format:
\{
    "evaluation" : "The current idea omits any concrete metrics or procedures, so it cannot yet demonstrate that the produced citation-hop dataset is correct or useful for verifying biomedical claims. Prior work such as ```<Paper corpusId="268248445">Piecing It All Together</Paper>``` shows that both expert-led quality audits and baseline system benchmarks are needed to answer “Have we solved the problem?”. Without similar expert agreement scores and model-based baselines, success cannot be measured.",
    "suggestions" : \{
        "We will sample automatically extracted citation chains and confirm that every cited paper is extant and accurately quoted. We will report the hallucination rate (false or malformed citations) and compare it to baseline large-language-model generation without our pipeline." : \{
            "278962365" : "<Quote corpusId="278962365">In a sample of 10 claims containing 211 citations, every work cited by Valsci was confirmed to be extant…</Quote>",
            ...
        \},
        ...
    \}
\}

IMPORTANT: Your output should strictly be of the following form:

\{
    "evaluation" : "an evaluation of the idea segment",
    "suggestions" : \{
        "suggested rewrite 1" : \{
            "corpusId1" : "<Quote corpusId="278962365">quote 1</Quote>",
            ...
        \},
        ...
    \}
\}

IMPORTANT: Your rewrites should only be rewrites of this problem statement, NOT the contribution or evaluation. Focus only on making this problem statement better.
IMPORTANT: The response should be entirely self-contained, meaning any quotes from prior literature should be directly quoted. Additionally the reader should understand the answer without knowing the question.
IMPORTANT: All references of papers should include the first few words of the paper title and should be of the following form: ```<Paper corpusId="268248445">PaperWeaver</Paper>```
IMPORTANT: Your assessment should be in two sections. The first should be an evaluation, which should be 3 sentences long. And the second should be dictionary of idea suggestions, which should be a direct rewrite of the selected segment along with a quote from related literature and a short explanation to support the suggestion.
IMPORTANT: The response should be as useful as possible to the researcher, so define any jargon by inferring the researcher's experience based on the idea they wrote.
IMPORTANT: The suggestions should be based one what you think the researcher would be most interested in given their original idea.
IMPORTANT: The structure of the output should be self-contained and not enumerated.\par}

\medskip

\subsubsection{Prompt to construct a bipartite graph}
\label{ap-bipartite-graph}

{\ttfamily\small\obeylines You are helping a researcher assess their research idea. To do this, you will be helping the researcher understand what the contributions are of current literature and what future work and limitations have been identified by current literature. This will ultimately be used to help make the researcher's idea more concrete and novel.

Ultimately, the researcher wants to refine their research idea so they are addressing an exciting, existing limitation that has not been addressed yet or work on a project that has a lot of contributions which is one that the research community is really excited about addressing.

Given a list of limitations and future work described in current, related research papers and a list of contributions of these papers, you will note whether a contribution directly addresses a limitation or future work.

Decide the match between contribution and future work or limitation with deep care and wisdom, so that they are the most meaningful to the researcher who wants to understand what future work or limitations identified by papers related to their idea has been addressed by current contributions of these papers.

Here is a list of papers with their limitations and future works: \{LIMITATIONS\}

Here is a list of papers with their current contributions: \{CONTRIBUTIONS\}

Here is the researcher's idea: \{IDEA\}

IMPORTANT: Make sure ALL papers appear in your output. Do not miss any papers.

The output will be a JSON where the key is the name of the limitation or future work and the value is a dictionary where the key is "addressed\_by" and the value is a list of paper titles and contributions that address the future work mapped to short, self-contained explanations. If no papers address the future work or limitation, the "addressed\_by" key should be mapped to an empty list.

Here is an example of the output structure:
\{
  "(paper title X) Limitation X": 
    "addressed\_by": [
      "(paper1 title) paper1 contribution" : "short explanation", 
      "(paper2 title) paper2 contribution" : "short explanation"
      ],
  "(paper title Y) Future work Y": 
    "addressed\_by": [
      "(paper3 title) paper3 contribution": "short explanation", 
      "(paper4 title) paper4 contribution": "short explanation"
      ],
  ...
\}\par}

\medskip

\subsubsection{Prompt to evaluate a solution facet}
\label{ap-sol-eval}

{\ttfamily\small\obeylines You will be executing two tasks: you will evaluate the researcher's proposed solution in their research idea and you will provide literature-grounded, concrete suggestions for how they can improve their proposed solution/research contribution.

This is their entire idea: \{IDEA\}

This is the part they need to refine: \{IDEA\_SEGMENT\}

IMPORTANT: This is additional context that the researcher wants you to know: \{ADDITIONAL\_CONTEXT\}. If there is no context, ignore this.

To help the researcher refine this part of their idea, you will rely on the following information. This dictionary maps current limitations and future work in related literature to contributions of current literature.

\{LIM\_FW\_CONT\_MATCHING\}

This is a list of all of the current contributions of the selected papers: \{CONTRIBUTIONS\}

For your evaluation, you want to make sure that the researcher is answering the following broad questions in their proposed solution: 1. How do current solutions fail at solving the problem the researcher has identified? 2. Does the researcher's solution better address the problem?

To formally evaluate the novelty of the proposed solution, you will do the following: The proposed solution is novel if it addresses future work or limitations in the set of literature that has not yet been addressed. Given this list of current limitations and future work described in related research papers and how current contributions of some papers address these limitations or future work, you are tasked with deciding if the researcher's proposed solution addresses a limitation or future work in this dictionary. If the proposed solution addressess a limitation or future work that has not been addressed by any other work, then it is novel. If the researcher's contribution matches an existing contribution, then it is not novel. Your evaluation should be no more than 4 sentences, and should include excerpts from and references to as many related papers as possible provide the best assessment of the paper. The references of papers should include the first few words of the paper title and should be of the following form: ```<Paper corpusId="268248445">PaperWeaver</Paper>```.

After the assessment, you will provide 1-3 concrete suggestions for how the current formulation of the researcher's solution or contribution can be more novel, if applicable. The suggested rewrite should be something that the researcher can directly replace their current idea segment with. Be specific, but don't make up any names or numbers such as intended results or system or benchmark or dataset names. More formally, rewrite the researcher's idea so that the questions below are clearly answered. Then, include excerpts from relevant literature that could be used to improve the contribution formulation. In particular, quote the future work and limitations described in current work to affirm that the suggestion is novel, or quote any very related literature that supports the suggestion. If at any point, any of the following questions are currently inadequately answered by the researcher, then you should use the literature provided above and your knowledge of the researcher's idea to ensure that your suggested rewrite answers all of these questions. Here are the questions:
1. Does the solution accurately address the problem?
2. Will the solution succeed?
3. Is the solution novel?

IMPORTANT: Here is an example of the output format:
\{
    "evaluation" : "The current idea omits any concrete metrics or procedures, so it cannot yet demonstrate that the produced citation-hop dataset is correct or useful for verifying biomedical claims. Prior work such as ```<Paper corpusId="268248445">Piecing It All Together</Paper>``` shows that both expert-led quality audits and baseline system benchmarks are needed to answer “Have we solved the problem?”. Without similar expert agreement scores and model-based baselines, success cannot be measured.",
    "suggestions" : \{
        "We will sample automatically extracted citation chains and confirm that every cited paper is extant and accurately quoted. We will report the hallucination rate (false or malformed citations) and compare it to baseline large-language-model generation without our pipeline." : \{
            "278962365" : "<Quote corpusId="278962365">In a sample of 10 claims containing 211 citations, every work cited by Valsci was confirmed to be extant…</Quote>",
            ...
        \},
        ...
    \}
\}

IMPORTANT: Your output should strictly be of the following form:

\{
    "evaluation" : "an evaluation of the idea segment",
    "suggestions" : \{
        "suggested rewrite 1" : \{
            "corpusId1" : "<Quote corpusId="278962365">quote 1</Quote>",
            ...
        \},
        ...
    \}
\}

IMPORTANT: Your rewrites should only be rewrites of this contribution statement, NOT the problem or evaluation. Focus only on making this contribution statement better.
IMPORTANT: The response should be entirely self-contained, meaning any quotes from prior literature should be directly quoted. Additionally the reader should understand the answer without knowing the question.
IMPORTANT: All citations of papers should include the first few words of the paper title and should be of the following form: ```<Paper corpusId="268248445">PaperWeaver</Paper>```
IMPORTANT: Your assessment should be in two sections. The first should be an evaluation, which should be 3 sentences long. And the second should be dictionary of idea suggestions, which should be a direct rewrite of the selected segment along with a quote from related literature and a short explanation to support the suggestion.
IMPORTANT: The response should be as useful as possible to the researcher, so define any jargon by inferring the researcher's experience based on the idea they wrote.
IMPORTANT: The suggestions should be based one what you think the researcher would be most interested in given their original idea.
IMPORTANT: The structure of the output should be self-contained and not enumerated.\par}

\medskip

\subsubsection{Prompt to evaluate an evaluation facet}
\label{ap-eval-eval}

{\ttfamily\small\obeylines You will be executing two tasks: you will evaluate the researcher's proposed evaluation in their research idea and you will provide literature-grounded, concrete suggestions for how they can improve their proposed evaluation.
This is their entire idea: \{IDEA\}
This is the part they need to refine: \{IDEA\_SEGMENT\}
IMPORTANT: This is additional context that the researcher wants you to know: \{ADDITIONAL\_CONTEXT\}. If there is no context, ignore this.

For your evaluation, you want to make sure that the researcher is answering the following broad question in their proposed evaluation: How would they know they've solved the problem?

These are how evaluations of current, successful papers show how a solution solves a problem:
\{PROL\_SOL\_EVAL\}

To formally evaluate the proposed evaluation, you will do the following: given the problem and solution statement of the idea, the evaluation, and the method, does the evaluation and method correctly ensure that the solution is addressing the problem? Given a list of problem statements, solution statements, methods, and evaluations of related research papers, you will use these as few-shot examples to calibrate yourself to what a successful evaluation method would look like. Then you will decide whether the proposed evaluation will correctly evaluate whether the solution correctly addresses the problem. Your evaluation should be no more than 4 sentences, and should include excerpts from and references to as many related papers as possible provide the best assessment of the paper. The references of papers should include the first few words of the paper title and should be of the following form: ```<Paper corpusId="268248445">PaperWeaver</Paper>```.

After the assessment, you will provide 1-3 concrete suggestions for how the current formulation of the researcher's solution or contribution can be more novel, if applicable. The suggested rewrite should be something that the researcher can directly replace their current idea segment with. More formally, rewrite the researcher's idea so that the questions below are clearly answered. Then, include excerpts from relevant literature that could be used to improve the problem formulation. These excerpts should be direct quotes of related evaluations or supporting evidence. If at any point, any of the following questions are currently inadequately answered by the researcher, then you should use the literature provided above and your knowledge of the researcher's idea to ensure that your suggested rewrite answers all of these questions. Here are the questions:
1. Will the proposed evaluation ensure the solution accurately address the problem?
2. Will the evaluation succeed?

IMPORTANT: Your output should strictly be of the following form:

Evaluation: [3 sentences of evaluation, following citing paper formats]
Suggestion: \{ full rewrite1 with concrete sugestion: 
                \{ "evidence1": quoted evidence from a cited paper,
                   "paper1" : corpusId of paper1 
                  "explanation1": explanation of why this is a good rewrite grounded in the evidence,
                  ...
                \},
                full rewrite2 with concrete sugestion: 
                \{ "evidence2": quoted evidence from a cited paper,
                   "paper2" : corpusId of paper2
                  "explanation2": explanation of why this is a good rewrite grounded in the evidence,
                  ...
                \}
              ...
        \}

IMPORTANT: Your rewrites should only be rewrites of this evaluation statement, NOT the contribution or problem. Focus only on making this evaluation statement better.
IMPORTANT: The response should be entirely self-contained, meaning any quotes from prior literature should be directly quoted. Additionally the reader should understand the answer without knowing the question.
IMPORTANT: Ground all your responses in the literature provided. All quotes from papers should be of the following form: ```<Quote corpusId="268248445">PaperWeaver is good</Quote>```
IMPORTANT: All citations of papers should include the first few words of the paper title and should be of the following form: ```<Paper corpusId="268248445">PaperWeaver</Paper>```
IMPORTANT: Your assessment should be in two sections. The first should be an evaluation, which should be 3 sentences long. And the second should be dictionary of idea suggestions, which should be a direct rewrite of the selected segment along with a quote from related literature and a short explanation to support the suggestion.\par}
\medskip

\subsubsection{Prompt to identify affected facet segments}
\label{ap-id-affected-facet}

{\ttfamily\small\obeylines A researcher wants to edit part of their research idea. You are helping a researcher understand what other parts of their research idea they will need to edit.
Here are all of the segments of the research idea that could be edited: \{IDEA\_SEGMENTS\}
Here is the idea that was just edited: \{IDEA\_SEGMENT\}
Here is the assessment of that edit: \{NOVELTY\_ASSESSMENT\}
You are tasked with identifying which segments should also be edited. Return only a list of exactly quoted segments that should be edited.\par}

\medskip

\subsubsection{Prompt to identify missing segments}
\label{identify-missing}

{\ttfamily\small\obeylines You are tasked with helping a researcher understand where they should add information about a particular facet of their research idea that is currently missing. You are given a researcher's idea, and the following dictionary. The dictionary keys are three facets, problem, solution, and evaluation. For each of these three facets, the values are sentences in the original idea that describe this facet. If any of the values in the facet is empty, meaning there is no text yet describing this facet, then you will infer where text describing this facet should be placed. You will return the original seed idea text, with [FACET] in places where the particular missing facet should be included.
Here is the original idea: \{ORIGINAL\_IDEA\}
Here is the dictionary of facets: \{FACETS\}
IMPORTANT: If there is already text that states [Add evaluation here], [Add problem here], [Add contribution here], or [Add solution here], assume that this facet is accounted for and not missing. So do not include that that facet is missing.\par}

\medskip

\subsubsection{Prompt to identify relevant clusters}
\label{ap-identify-relevant-cluster}

{\ttfamily\small\obeylines A researcher is refining a part of their research idea. To refine their research idea, they will be reading papers in the following clusters of papers. You will be tasked with identifying which paper clusters are most relevant to the part of the idea that the researcher is refining.
Here is the part of the idea that the researcher is editing: \{IDEA\_SEGMENT\}
Here are the names of the paper clusters: \{CLUSTERS\}
IMPORTANT: Please return ONLY the exact cluster names, one per line, with no explanations or numbering. Select 1-3 clusters that most relate to the part of the idea.
Example output format:
Challenges in verifying health and medical claims
Citation integrity and provenance tracking\par}

\medskip

\subsubsection{Prompt to cluster by the contribution facet}
\label{ap-cluster-contribution}

{\ttfamily\small\obeylines You will be helping a researcher with a research idea understand the clusters of current contributions in the research space that this idea is in so that they can understand what the main contribution patterns are that current projects make. Given a list of passages of current contributions from related research papers, you will come up with meaningful clusters to describe groups of contributions in this research area. You should make sure that every passage is included in a cluster. Choose the clusters with deep care and wisdom, so that they are the most meaningful to the researcher who wants to understand what the current contribution patterns are in papers related to their idea.
Here is a list of corpusId and contribution passages: \{CONTRIBUTIONS\}
Here is the seed idea: \{SEED\}
IMPORTANT: There are \{NUM\_UNIQUE\_CORPUS\_IDS\} unique corpus IDs in the data below. Make sure ALL of them appear in your clustering output. Do not miss any corpus IDs.
The output will be a JSON where the key is the name of the cluster, typically a sentence that describes the contribution cluster, and the value is a list of just corpusIds that are in this cluster. Do not include the contribution passages in the output. Only include the corpusId.
Here is an example of the output structure:
\{
  "Novel algorithmic approaches and theoretical frameworks": ["corpusID1", "corpusID2"],
  "Empirical evaluation and benchmark improvements": ["corpusID3"]
  ...
\}\par}

Note: We use a similar prompt structure to cluster by other facet types.

\medskip

\subsubsection{Prompt to generate complete assessments}
\label{ap-gen-complete-assessment}

{\ttfamily\small\obeylines You will be giving a full assessment of this researcher's project proposal. The proposal should answer these questions: 
1. What is the problem? This should be of the form "We currently can't do Y" or "We want to do Y"
2. What is the specific problem we are trying to solve?
3. How do we know this is an actual problem?
4. Why should we bother solving this problem?
5. Does the solution accurately address the problem?
6. Will the solution succeed?
7. Is the solution novel?
8. Will the proposed evaluation ensure the solution accurately address the problem?
9. Will the evaluation succeed?
Here are previous assessments of parts of this idea.
\{PREVIOUS\_ASSESSMENTS\}
Here is the proposal: \{IDEA\}
Your assessment should be no more than 5 sentences. Be thorough.\par}

\section{Example of ideas from \system and the baseline from the user study}

\begin{table*}[p]
\centering
\footnotesize
\setlength{\tabcolsep}{4pt}
\renewcommand{\arraystretch}{1.05}
\caption{Side-by-side comparison of baseline ideas and their LitPivot counterparts. Diagonal entries are from the same participant.}
\Description{Three columns (Initial Idea, Baseline, LitPivot) and two rows detailing different research concepts. Row 1 compares paragraph-length baseline and LitPivot expansions for "Initial Idea 1," which proposes a tool to help users engage with and verify the quality of source texts in RAG QA systems. Row 2 compares paragraph-length baseline and LitPivot expansions for "Initial Idea 2," which proposes an interactive semi-formal proof navigator to verify scientific claims across research papers using logical propositions. Note: The table contains text in the Baseline and LitPivot columns demonstrating how each participatn expands on the initial ideas.}
\label{tab:baseline-vs-litpivot}
\begin{tabular}{|p{.20\textwidth}|p{.38\textwidth}|p{.38\textwidth}|}
\hline
& \textbf{Baseline} & \textbf{LitPivot} \\
\hline

\textbf{Initial Idea 1:} Current RAG QA systems are touted as more interpretable and explainable because they point to existing sources. But sometimes these sources aren't even relevant to the generation. We propose a tool that helps users engage more with source texts and better discern the quality of source texts. &
Recent research has highlighted a critical gap in RAG QA systems where retrieved documents that appear relevant based on surface-level similarity, don't actually support the specific claims being made in the generation.  While RAG systems theoretically provide transparency by surfacing source documents, the granularity of attribution is often too coarse (typically at the document or sentence level) to be truly useful for verification, as demonstrated in works on fine-grained attribution. The mismatch between retrieved sources and generated content not only undermines the system's interpretability claims but also potentially misleads users who trust the system's ability to ground its responses in reliable sources. We propose a tool that combines fine-grained attribution techniques with an interactive interface, where users can click on specific claims in the generated text to see the exact supporting quotes from source documents highlighted in context. The tool also employs a traffic light system to visually indicate the strength of attribution for each claim (green for direct quotes/strong support, yellow for partial/indirect support, red for unsupported claims), helping users quickly identify potential issues. We can evaluate our tool through a multi-faceted approach combining: (1) precision/recall metrics comparing system-identified attributions to human-annotated gold standards, (2) semantic similarity scores between generated claims and source text segments using established embedding models, (3) user studies measuring verification time and accuracy with and without the tool. &
Current RAG QA systems are touted as more interpretable and explainable because they point to existing sources. However, these systems still tend to fail for certain tasks, e.g. they can hallucinate information, miss key references, and find wrong or unrelated sources. Existing literature focus on attribution (and how to use them for improving QA for LLMs, but not how to engage with these attributions, nor how to help users discern the quality of these attributions. Enabling users to do so empowers users to be more diligent in checking the answers and sources' quality and improves the trustworthiness and transparency of such RAG QA systems. We propose a tool that helps users engage more with source texts and better discern the quality of source texts. We introduce a voice- and pointer-driven RAG interface in which (i) users can issue spoken follow-ups such as show me the sentence that supports that claim or read the most reliable paragraph aloud, and (ii) every cited span is accompanied by an automatically generated reliability badge that aggregates contradiction, sufficiency and source-authority scores computed with CiteEval-AUTO and AutoAIS.  The system also supports LAQuer-style localized queries so that users can drill down into sub-sentences, with the interface highlighting the exact supporting tokens.   Our tool provides 1) a novel way for users to interact with these sources through multimodal interaction paradigms including source navigation (via attribution) and voice interaction, rather than treating RAG as a blackbox component in the QA pipeline, and 2) a novel way for users to easily verify and discern source quality. We focus on textual content as the source, leaving multimodal content (images, videos, PDFs, voices) as future work. A controlled user study will measure whether the badge + drill-down workflow improves verification speed and accuracy compared to the baseline clickable-citation UI used in TruthReader. \\
\hline

\textbf{Initial Idea 2: } Existing methods for verifying scientific claims across research papers typically lack systematic checks for logical consistency. We want to develop an interactive semi-formal proof navigator that translates natural-language claims from research papers into structured logical propositions. &
Existing methods for verifying scientific claims across research papers typically lack systematic checks for logical consistency. We want to develop an interactive, two-way ``semi-formal proof navigation'' system that 1) translates natural-language claims from research papers into structured logical propositions using various formal logic representations for flexibility, 2) enables visualizing and interacting with the claims and their logical relationships. The system automatically highlights parts with problematic claims and suggests remediations, and enables the user to modify visualized logical claims where the results will be updated in the original content, and vice versa. &
Scientists currently lack automated tools that can detect when claims drawn from different papers form a logically inconsistent chain of reasoning. Existing scientific fact-checking resources focus on single-claim, single-paper links for example, SCIFACT targets abstract-level support or refutation, and CLAIM-BENCH evaluates claim-evidence reasoning within one paper leaving cross-paper logical consistency unchecked. Without such cross-document verification, meta-analyses and policy decisions risk propagating contradictions or unsupported citation leaps in an ever-growing literature, making a principled consistency checker both timely and necessary. We will build an interactive proof-navigation environment that (i) harvests claims and supporting sentences from multiple research papers, (ii) auto-formalises each statement into FOL while attaching citation metadata, (iii) dynamically chooses the most reliable symbolic solver per claim type, (iv) stitches the resulting formulas into a cross-paper proof graph, and (v) lets users step through, edit, and re-verify any node to expose contradictions or unsupported leaps. \\
\hline

\end{tabular}
\end{table*}

\clearpage

\section{Affected Facets feature}
\begin{figure}[h]
    \centering
    \includegraphics[width=0.7\columnwidth]{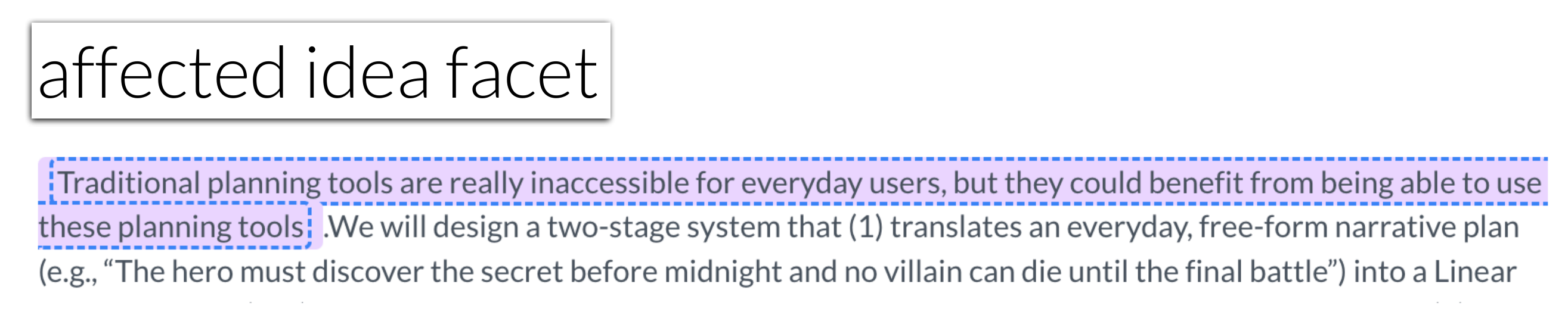}
    \caption{When a change to one idea facet might require editing another idea facet, \system highlights the corresponding affected idea facet in the text editor. Here, the problem statement facet is outlined.}
    \Description{An editor view with the label ``affected idea facet.'' The first sentence
    of the idea statement (``Traditional planning tools are really inaccessible\ldots'')
    is outlined with a dashed purple box, indicating the specific facet of the idea that
    a critique or suggestion will modify.}
    \label{fig:affected-facet}
\end{figure}

\section{Baseline}
\begin{figure}[h]
    \centering
    \includegraphics[width=\columnwidth]{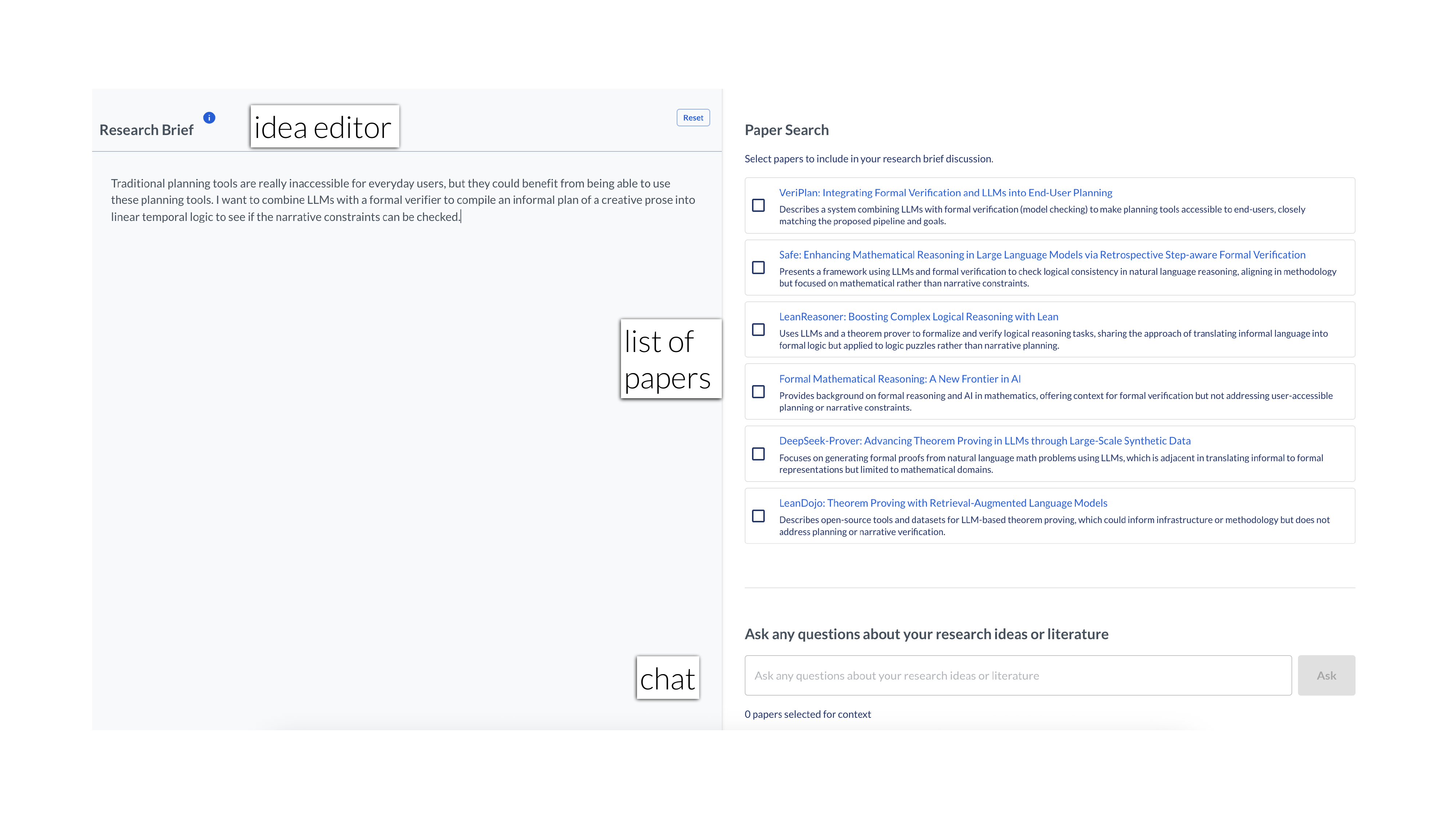}
    \caption{\textbf{Baseline interface.} The left pane contains an idea editor where researchers can write out their research idea. The right pane shows a list of retrieved papers. At the bottom, a chat box allows researchers to ask questions about their idea or the listed literature.}
    \Description{A baseline interface with three main components. On the left, an idea
    editor pane lets researchers write out their research idea in free text. On the
    right, a list of papers displays retrieved literature relevant to the idea, each
    with a short description. At the bottom, a chat box invites researchers to ask
    questions about their idea or the listed papers, supporting interactive exploration.}
    \label{fig:baseline}
\end{figure}

\section{Design Probe}
\label{design-probe}
\begin{figure}[h]
    \centering
    \includegraphics[width=0.5\columnwidth]{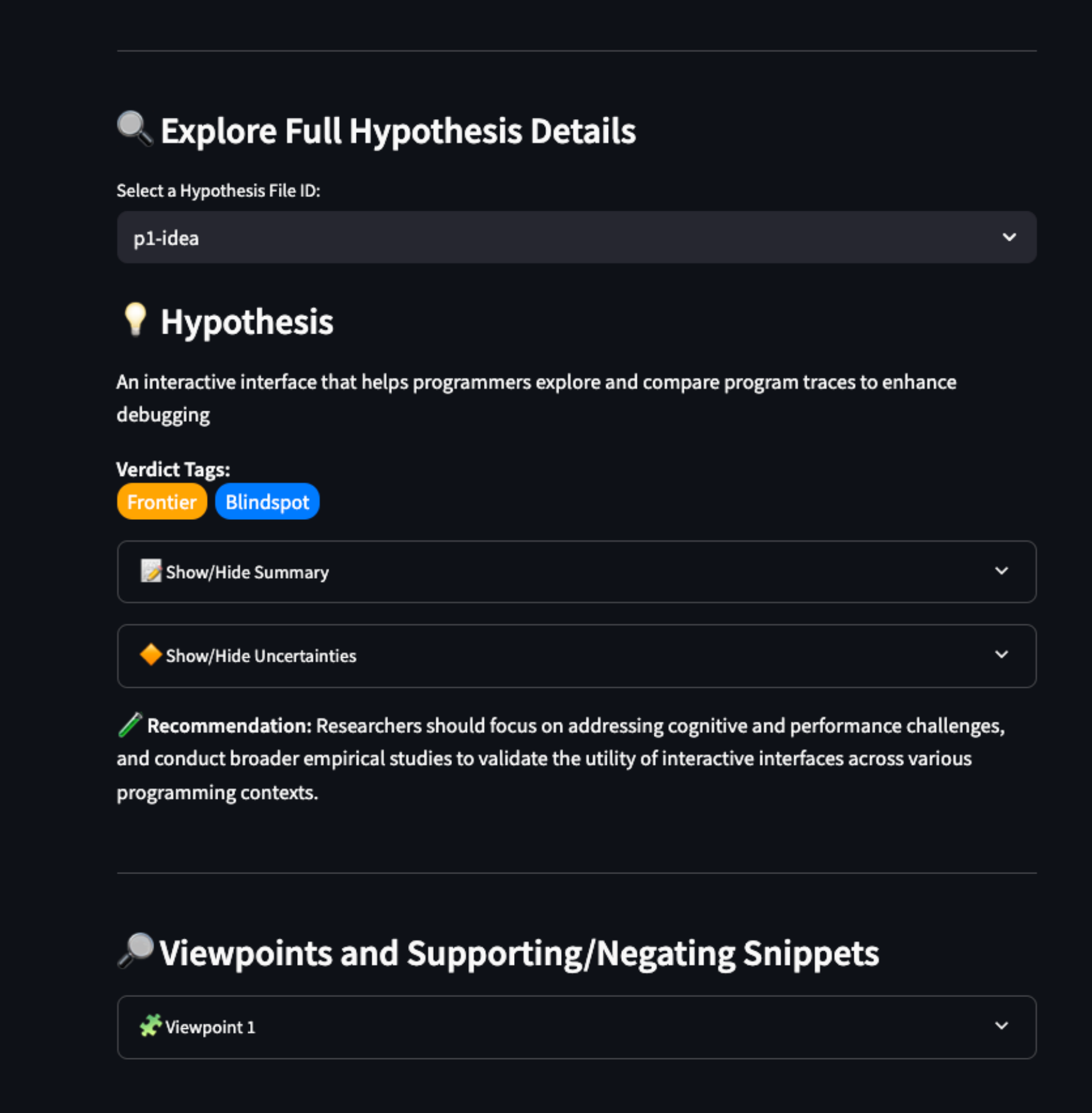}
    \caption{Design probe used in the formative study. This is an LLM-powered prototype. The research hypothesis is displayed at the top. Verdict tags are assigned below. There are recommendations, supporting and negating viewpoints that use related literature to ground feedback.}
    \Description{The design probe interface. At the top, the user can select a hypothesis
    file ID from a dropdown. The selected hypothesis is displayed with a description
    (``An interactive interface that helps programmers explore and compare program
    traces to enhance debugging'') and is tagged with the verdict tags ``Frontier'' and
    ``Blindspot.'' Buttons allow the user to expand or collapse a summary and view
    uncertainties. A green-highlighted recommendation suggests focusing on cognitive and
    performance challenges and conducting broader empirical validation. Below, a section
    titled ``Viewpoints and Supporting/Negating Snippets'' lists viewpoint entries that
    can be expanded to show related evidence from the literature.}
    \label{fig:design-probe}
\end{figure}


\end{document}